\documentclass[10pt,aps,prl,twocolumn,superscriptaddress]{revtex4-2}
\usepackage{amsfonts, mathtools, amsthm, bm}
\usepackage{braket}
\usepackage{xcolor}
\usepackage[breaklinks=true,colorlinks,allcolors=blue]{hyperref}
\usepackage{comment}
\usepackage{orcidlink}
\usepackage[normalem]{ulem}

\newtheorem{theorem}{Theorem}

\newcommand{\diag}{\text{diag}}

\newcommand{\trace}[1]{\text{tr}[#1]}
\newcommand{\norm}[1]{\left\lVert#1\right\rVert}

\newcommand{\eqlabel}[1]{equation~\eqref{#1}}
\newcommand{\figlabel}[1]{Fig.~\ref{#1}}

\newcommand{\theolabel}[1]{Theorem~\ref{#1}}
\newcommand{\hc}{\text{h.c.}}
\newcommand{\sm}{Supplementary Information} 
\newcommand{\I}[1]{\Im{[#1]}}
\newcommand{\R}[1]{\Re{[#1]}}

\begin{document}

\title{Cumulative locality governs collective quantum charging}

\author{Sebastián V. Romero\orcidlink{0000-0002-4675-4452}}
\email{sebastian.v.romero@csic.es}
\affiliation{Quantum Advanced Research Center (QuARC), CSIC, 28049 Madrid, Spain}
\affiliation{Instituto de Ciencia de Materiales de Madrid (ICMM), CSIC, 28049 Madrid, Spain}
\affiliation{\mbox{Departamento de Física Teórica de la Materia Condensada, Universidad Autónoma de Madrid, 28049 Madrid, Spain}}

\author{Xi Chen\orcidlink{0000-0003-4221-4288}}
\email{xi.chen@csic.es}
\affiliation{Quantum Advanced Research Center (QuARC), CSIC, 28049 Madrid, Spain}
\affiliation{Instituto de Ciencia de Materiales de Madrid (ICMM), CSIC, 28049 Madrid, Spain}

\author{Yue Ban\orcidlink{0000-0003-1764-4470}}
\email{yue.ban@csic.es}
\affiliation{Quantum Advanced Research Center (QuARC), CSIC, 28049 Madrid, Spain}
\affiliation{Instituto de Ciencia de Materiales de Madrid (ICMM), CSIC, 28049 Madrid, Spain}
\date{\today}

\begin{abstract}
    Quantum batteries can outperform their classical counterparts through collective quantum effects, yet the conditions enabling such advantage remain unclear. Using a generic class of free-fermionic chargers we prove that global operators are insufficient for scalable superextensive charging and that a cumulative hierarchy of many-body operators is required, with the structure of the battery determining the degree of advantage. This is obtained by relating the charging power to the norm of the commutator between the battery and charger Hamiltonians and deriving an upper bound only capturing noncommuting terms.
    Moreover, we revisit the concept of participation number and propose an effective definition that accounts for the relative strength of the contributing terms, correlating it with the charging power scaling. Not limited to quadratic Hamiltonians, our results establish cumulative locality as a key resource for quantum charging advantage.
\end{abstract}

\maketitle

Quantum batteries (QBs) exploit quantum coherence and many-body effects to store and release energy, offering potential advantages over classical devices. A central goal is to understand the mechanisms behind charging advantage. Collective quantum effects can enhance charging performance, with global operations boosting power beyond local strategies, although the resulting advantage is bounded by the interaction order~\cite{campaioli2017enhancing}. Further studies linked charging power to physical quantities such as Fisher information and multipartite entanglement~\cite{juliafarre2020bounds}. More recently, it has been shown that global operations can provide at most an extensive speedup over classical protocols~\cite{gyhm2022quantum}.%

Despite recent progress, the conditions for superextensive scaling of charging power remain elusive, as it is unclear whether global operations can guarantee a scalable advantage. In this work, we identify the conditions required for achieving superextensive scaling in the charging power, using free-fermionic chargers as paradigmatic examples. Although these models are quadratic in fermions, their spin-$1/2$ representation contains higher-order strings, providing a minimal setting that combines analytical tractability with many-body contributions. 
We show that global operators are insufficient for scalable superextensive scaling. 
Instead, the charger must exhibit \emph{cumulative locality}, i.e., a hierarchy of operator families spanning over increasing localities with the maximum locality $n_L$ growing with system size
(represented as piled plates in~\figlabel{fig:scheme}), which holds even under the presence of locality-decaying interactions. 
This is captured by a derived upper bound on the charging power corresponding to the norm of the commutator between the battery and charger Hamiltonians, whose scaling is determined by the noncommuting terms present. 
To make this connection explicit, we propose the Hill diversity index~\cite{hill1973diversity} as effective participation number and correlate it with the charging power scalings. 
Not limited to quadratic Hamiltonians, we also validate our findings by means of the Sachdev-Ye-Kitaev (SYK) model~\cite{rossini2020quantum,divi2025syk} and a global charger. Altogether, this work characterizes the prerequisites for a scalable charging advantage, providing a unified framework for the design of QBs.
\begin{figure*}[!tb]
    \centering
    \includegraphics[width=\linewidth]{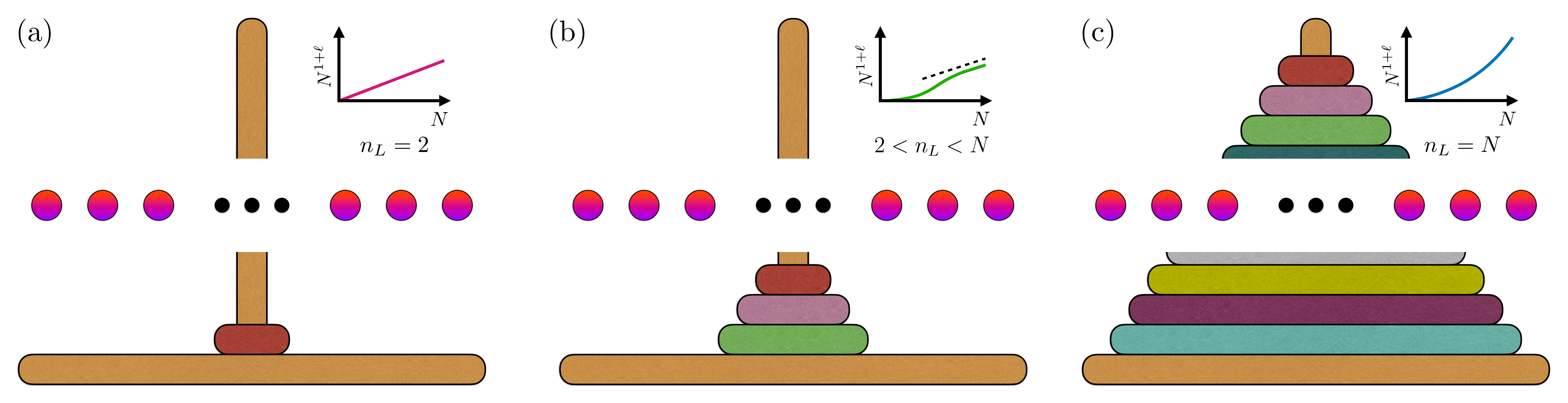}\vspace{-1mm}%
    \caption{\textbf{Schematic illustration of cumulative locality.} A set of spins (spheres) interacts through families of many-body operators generated by the charging Hamiltonian. The diameter of Hanoi-tower plates represents the operator locality, while the stacked plates depict the cumulative hierarchy up to the maximum locality $n_L$. 
    \textbf{a,} For $n_L=2$, only local and two-body terms are present, leading to extensive scaling. 
    \textbf{b,} For a fixed $2<n_L<N$, the locality hierarchy is truncated, leading to a crossover around $n_L \sim N$ and asymptotically extensive scaling for $n_L \ll N$. 
    \textbf{c,} For $n_L=N$, the system reaches the fully global regime (all operators in \eqlabel{eq:charger_jwt} are present), enabling superextensive scaling.%
    }\label{fig:scheme}\vspace{-3mm}
\end{figure*}

We consider a broad class of QBs described by a local battery $H_\text{b}=\sum_{i=1}^N h_i$, with $h_i = \alpha_i I_i + \beta_i \sigma^x_i + \gamma_i \sigma^y_i + \delta_i \sigma^z_i$, and a quadratic charger Hamiltonian~\cite{lieb1961two}%
\begin{equation}\label{eq:quadratic}
    H_\text{c}(t) = \sum_{k,l=1}^N \!\left[ c^\dagger_k A_{kl}(t) c_l + \frac{1}{2}\left( c^\dagger_k B_{kl}(t) c^\dagger_l + \hc \right) \right],
\end{equation}%
where $c^\dagger_i$ ($c_i$) are fermionic creation (annihilation) operators, with $A$ and $B$ Hermitian and skew-symmetric matrices, respectively. 
Applying the Jordan-Wigner transformations $\sigma^+_i = -c_i\prod_{j<i}(1-2c^\dagger_jc_j)$ and $\sigma^z_i = 1-2c^\dagger_ic_i$, together with $\{c^\dagger_i,c_j\}=\delta_{ij}$ and $\{c_i,c_j\}=0$, the charger can be written in the spin-$1/2$ representation as 
\begin{widetext}\vspace{-3mm}
\begin{equation}\label{eq:charger_jwt}
\begin{split}
    H_\text{c}(t) &= \frac{1}{2}\sum_{k=1}^N A_{kk}(t)(1-\sigma^z_k) +\frac{1}{2}\sum_{k<l} \Big[ (\R{A_{kl}(t)} + \R{B_{kl}(t)})\sigma^x_k S_{kl} \sigma^x_l + (\R{A_{kl}(t)} - \R{B_{kl}(t)})\sigma^y_k S_{kl} \sigma^y_l  \\
    &\qquad\qquad -(\I{A_{kl}(t)} - \I{B_{kl}(t)})\sigma^x_k S_{kl} \sigma^y_l + (\I{A_{kl}(t)} + \I{B_{kl}(t)})\sigma^y_k S_{kl} \sigma^x_l\Big].
\end{split}
\end{equation}\vspace{-3mm}
\end{widetext}
with $S_{ij}= \prod_{k=i+1}^{j-1} \sigma^z_k$ the string operator. 
The terms appearing in~\eqlabel{eq:charger_jwt} have locality $l-k+1$, ranging from local to $N$-body operators. Cumulative locality is therefore defined as the buildup of operator families across different localities up to the maximum locality denoted by $n_L$. %

Starting from the initial state $\rho(0)$ and taking $\hbar\equiv 1$, the system evolves as $\dot{\rho}(t) = -i[H_\text{c}(t), \rho(t)]$. The energy injected reads as $E(t) = \trace{(\rho(t)-\rho(0))H_\text{b}}$ with instantaneous power $P(t) = %\dot{E}(t) =
\trace{\dot{\rho}(t)H_\text{b}} = -i\trace{\rho(t)[H_\text{b}, H_\text{c}(t)]}$. 
Its scaling with system size, $P(t) \sim N^{1+\ell}$,  characterizes the charging performance: $\ell =0$ corresponds to extensive scaling, while $0<\ell\le1$ to superextensive, only genuinely reachable by quantum means~\cite{campaioli2017enhancing, gyhm2022quantum}. This scaling is bounded by (see Methods)%
\begin{equation}\label{eq:bounds}
    |P(t)| \le \norm{ [H_\text{b}, H_\text{c}(t)] } \le 2 \norm{H_\text{b}} \norm{H_\text{c}(t)},
\end{equation}
where $\norm{\cdot}$ denotes the operator norm and the rightmost inequality follows from the triangle inequality. Although the right-hand side bound is widely used, it can be loose when many contributions commute. 
To exploit the commutative structure, we decompose $[H_\text{b}, H_\text{c}(t)]$ into four operator families, with $\sum_{i=1}^N [\beta_i\sigma^x_i + \gamma_i\sigma^y_i, H_\text{c}(t)] = C_1 + C_2 + C_3$ and $\sum_{i=1}^N [\delta_i\sigma^z_i, H_\text{c}(t)] = C_4$. Here, $C_1$ contains local $\sigma^{(x,y)}_k$ contributions, while $C_2$ contains $\sigma^\pm_k S_{kl} \sigma^\pm_l S_{lm} \sigma^\pm_m$ terms. The remaining  $C_{(3,4)}$ families encapsulate noncommuting summands with respect to the endpoints present in the many-body operators of~\eqlabel{eq:charger_jwt}, with $C_3$ ($C_4$) including $S_{k-1,l}\sigma^\pm_l$ and $\sigma^\pm_k S_{k,l+1}$ ($\sigma^\pm_k S_{kl} \sigma^\pm_l$) terms. 
This decomposition tracks how different locality sectors contribute to the commutator. Using Hamiltonians with terms up to maximum locality $n_L$, we study how cumulative locality affects charging-power scaling with system size.%

We now present our main result (proved in the \sm), which bounds $[H_\text{b}, H_\text{c}(t)]$, thus on $P(t)$, by resolving the contributions from different operator families. Hereinafter, we omit the explicit time dependence of $H_\text{c}(t)$.%. 
\begin{theorem}\label{th:result}
    Let $H_\textup{b}=\sum_{i=1}^N h_i$ with $h_i = \alpha_i I_i + \beta_i \sigma^x_i + \gamma_i \sigma^y_i + \delta_i \sigma^z_i$ and $\alpha_i,\beta_i,\gamma_i,\delta_i\in\mathbb{R}$ $\forall i$ and $H_\textup{c}$ as in~\eqlabel{eq:quadratic}. The instantaneous power $P(t)$ satisfies%
\begin{equation}\label{eq:final}
     |P(t)|\le 
    \norm{[H_\textup{b}, H_\textup{c}]}  \le \mathcal{U}_{n_L}(N),
\end{equation}\vspace{-3mm}
with 
\begin{widetext}\vspace{-3mm}
    \begin{equation}\label{eq:final_bound}
    \mathcal{U}_{n_L}(N)= 4b_\perp (n_L-2)\sqrt{N(N-2)} + 
    \sqrt{2}\sigma_L \left[ \Lambda_1 N +2(\Lambda_3 + \Lambda_4)\left[ \sum_{k=1}^{n_L-1} k^{1/2} + (N-n_L)\sqrt{n_L-1} \right] \right] ,
\end{equation}
For the fully global case $n_L=N$, the bound takes the asymptotic form
\begin{equation}\label{eq:final_limit}
    \mathcal{U}_{N}(N)=
    4b_\perp N^2 + \left[ \frac{8}{3}(\Lambda_3 + \Lambda_4) -12b_\perp \right] N + 2 \Lambda_1 N^{1/2} + 2\left( 3b_\perp - \Lambda_3 - \Lambda_4 \right) + \mathcal{O}[N^{-1/2}] ,
\end{equation}\vspace{-3mm}
    \end{widetext}%
\end{theorem}%
\noindent 
with $\sigma^2_L = 2N/[N+(n_L-1)(2N-n_L)]$ the normalization factor of $H_\text{c}$ (Methods), $\Lambda_i$ the largest absolute Pauli coefficient of $C_i$ in absolute value, and $b_\perp^2 = \max_{2\le l\le N-1} (\beta_l^2 + \gamma_l^2)$. 
The bound is obtained by reducing the norm of to a sum of the norms of the four operator families and tracking their scaling with system size. 
In particular, while the operator norm of $C_1$ can be evaluated exactly, the norm of $C_2$ is bounded by the charger Hamiltonian renormalization and the norms of $C_{(3,4)}$ are bounded by grouping terms into pairwise anticommuting subfamilies, whose squares are proportional to the identity.
The resulting bounds are expressed in terms of Pauli coefficients, which are proportional to the $H_\text{b}$ coefficients and the entries of $A$ and $B$.
Their scaling is therefore determined by the number of operators present in the charger, subject to the locality cutoff $n_L$ and the normalization of $H_\text{c}$. %
Equations~\eqref{eq:final_bound} and~\eqref{eq:final_limit}
provide a natural classification according to the $C_i$ families: while $C_{(1,3,4)}$ yield at most extensive scaling, the $C_2$ family, containing up to $8\binom{N}{3}$ terms, gives rise to the quadratic term in the asymptotic bound, where a linearly-growing locality cutoff (i.e., $n_L\propto N$) is mandatory. 
More generally, for $n_L \propto N^\nu$ with $0\leq\nu\leq1$, the leading
contribution scales as $\mathcal{U}_{n_L}(N)\sim N^{1+\nu}$ (Methods),
directly linking the growth of the locality hierarchy to the charging power scaling exponent.%

\begin{figure}[!tb]
    \centering
    \includegraphics[width=\linewidth]{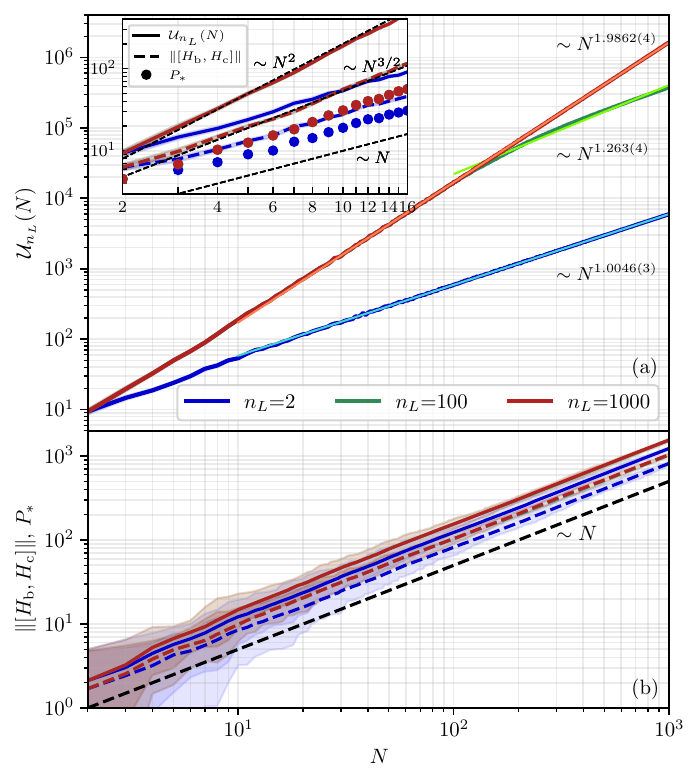}\vspace{-2mm}%
    \caption{\textbf{Scaling of the charging power bound.} 
    \textbf{a,} The bound $\mathcal{U}_{n_L}(N)$ in~\eqlabel{eq:final_bound} for $n_L=\{2,100,1000\}$. 
    Data points are fitted to $f(N)=aN^b$, excluding $N\le100$ for $n_L=100$ and $N\le10$ otherwise. The $n_L=100$ and $n_L=1000$ curves overlap for $N\le100$. 
    The inset compares the bound with the exact commutator norm $\norm{[H_\text{b},H_\text{c}]}$ and the maximum charging power $P_*$ for $N\in[2,16]$. 
    \textbf{b}, Exact commutator norms (solid) and $P_*$ (dashed) for the quadratic battery subclass, with $n_L=\{2,100,1000\}$. All cases remain extensive despite the increasing locality cutoff. All results (both \textbf{a} and \textbf{b}) are averaged over $32$ disorder realizations, with shaded areas indicating the max-min range.
    }\label{fig:results}%
\end{figure}

To assess the tightness of~\eqlabel{eq:final}, we compare the upper bounds with exact operator norms for different system sizes and locality cutoffs $n_L$. 
We retain localities up to $n_L$ by setting $A_{kl}(n_L) = a_{kl}$ and $B_{kl}(n_L) = b_{kl}$ if $|k-l|< n_L$ and zero otherwise. The real and imaginary parts of $a_{kl}$ and $b_{kl}$, together with the coefficients of $H_\text{b}$, are normally distributed, with the charger couplings renormalized for a well-defined thermodynamic limit. We also compute the maximum charging powers $P_*= \max_t |P(t)|$, using the ground state of $H_\text{b}$ as the initial state. 
Figure~\ref{fig:results}a shows the upper bound of $\norm{[H_\text{b}, H_\text{c}]}$ for $N\in[2,1000]$ and $n_L\in\{2,100, 1000 \}$ along with the exact norms and $P_*$ for $N\in[2,16]$. 
For $n_L=2$, the bound scales extensively, wheareas for $n_L=1000$, it approaches the $N^2$ scaling. 
The exact norms and $P_*$ shows a $\sim N^{3/2}$ scaling, suggesting that our upper bound may not be asymptotically tight. For $n_L=100$, the cutoff is inactive for $N \leq 100$, so the charger is effectively fully global, yielding an initial superextensive scaling.  For $N>100$, the truncated locality hierarchy discards higher-order operator families that results in a crossover to extensive scaling, confirming that superextensive scaling requires a locality hierarchy that grows with $N$.%

A particular subclass arises when $H_\text{b}$ is itself quadratic, thus $\beta_i,\gamma_i=0$. 
In this case, $[H_\text{b},H_\text{c}]=C_4$ remains quadratic and its norm can be computed in polynomial time by casting it in a block form (\sm). Defining $D=\diag(\delta_1,\dots,\delta_N)$ and the Nambu spinor $\Psi = [c_1, \dots, c_N, c^\dagger_1, \dots, c^\dagger_N]^\text{T}$, we can write 
\begin{equation}\label{eq:quadratic_c}
    [H_\text{b}, H_\text{c}] = C_4 = \frac{1}{2} \Psi^\dagger \mathcal{C} \Psi \quad\text{ }\text{ } \text{with }\mathcal{C} = \begin{bmatrix} \mathcal{A} & \mathcal{B} \\ \mathcal{B}^* & -\mathcal{A}^\text{T} \end{bmatrix},
\end{equation}
where $\mathcal{A} = 2[A,D]$ and $\mathcal{B} = -2\{B,D\}$. In~\figlabel{fig:results}b, we compare the exact operator norms with the maximum charging powers for normally distributed $\delta_i$. 
Reamarkably, regardless of the locality cutoff $n_L$ set, the charging powers scale extensively, in agreement with~\theolabel{th:result}. 
The $\ell=0$ exponent follows the battery Hamiltonian choice, which only contains $\sigma^z_j$ terms and therefore commutes with the strings $[\sigma^z_j,S_{kl}]=0$. 
As a consequence, although cumulative locality is present, the string-mediated contributions responsible for superextensive scaling are absent from $[H_\text{b},H_\text{c}]$. 
This shows that the battery structure also determines which operator
families contribute to the charging-power bound in~\eqlabel{eq:final}.%

Having established the correspondence between the power law scaling and $n_L$, we revisit the notion of participation number. Defined as the number of operators in which each spin participates~\cite{campaioli2017enhancing}, its precise role in charging advantages remains under debate~\cite{gyhm2022quantum}. Since only charger terms that do not commute with the battery can inject energy, we introduce a commutator-resolved participation number, constructed directly from the local commutators $[h_k,H_\text{c}]$. 
Consequently, we write as Pauli decomposition for the $k$th cell $[h_k,H_\text{c}] = \sum_{s\in S^{(k)}} c^{(k)}_s P_s$, with $S^{(k)}$ the set of Pauli strings in the commutator and $c^{(k)}_s$ their coefficients. Assigning the normalized weights $p^{(k)}_s = |c^{(k)}_s|/\sum_{s'\in S^{(k)}} |c^{(k)}_{s'}|$, we propose the effective participation number through the Hill diversity index $N^{(k)}_1 = \exp\!\big[-\sum_{s\in S^{(k)}} p^{(k)}_s \log p^{(k)}_s \big]$, which quantifies 
the effective number of strings contributing to energy injection while accounting for their relative strengths (Methods). This provides a direct bridge between the charger structure and the charging power scaling, which is sharper than simply counting going beyond a simple count of noncommuting terms.%
\begin{figure}[!tb]
    \centering
    \includegraphics[width=\linewidth]{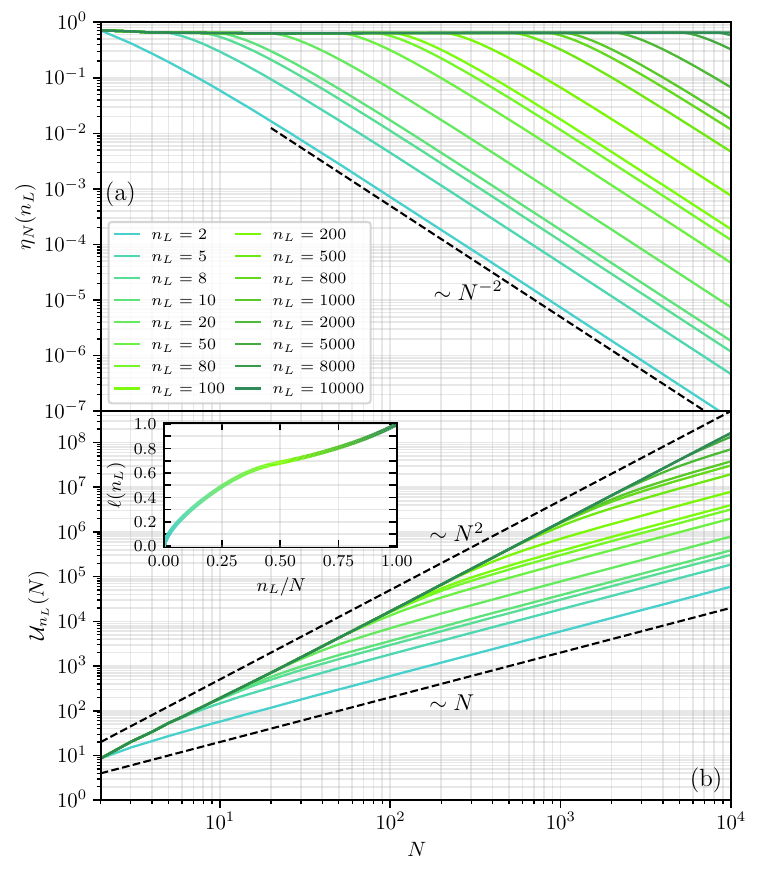}\vspace{-2mm}%
    \caption{\textbf{Link between participation ratio and charging power scaling.}  \textbf{a,} Participation ratio $\eta$ as a function of $N$ for locality cutoffs $n_L$ ranging from $n_L=2$ to $10000$. 
    \textbf{b,} Corresponding upper bounds $\mathcal{U}_{n_L}(N)$ for the same locality cutoffs. 
    The inset shows the fitted scaling exponent $\ell(n_L)$ obtained by fitting $\mathcal{U}_{n_L}(N)$ to $aN^{1+\ell}$ using data with $N>n_L$, while for the fully global endpoint $n_L=N=10000$, all system sizes are included in the fit. All results are averaged over a hundred disorder realizations.}\label{fig:participation}%
\end{figure}%

To relate the participation number to the power scaling, we define the participation ratio $\eta= N_1/N_{\max}$ with $N_1 = \sum_{k=1}^N N^{(k)}_1$ and $N_{\max} = 2N(N+2)(2N-1)/3\sim\mathcal{O}[N^3]$, which corresponds to the total number of contributing terms in the fully global case. 
In~\figlabel{fig:participation}, we compare $\eta$ and the power scaling exponent $\ell$ ranging from the two-body $(n_L=2)$ to the fully global regime $(n_L=10000)$. 
Both quantities are computed from random realizations of $A(n_L)$, $B(n_L)$ 
and $H_\text{b}$ for system sizes up to $N\in[2,10000]$. 
The scaling exponents are extracted from fits of $\mathcal{U}_{n_L}(N)$ for $N\ge n_L$, 
except for $n_L=N=10000$, where all system sizes are included. 
As shown in~\figlabel{fig:participation}a, the participation ratio $\eta$ displays a plateau near $\eta_\text{crit}\approx 0.6529$ in the fully global regime $n_L\simeq N$ (\sm), where the locality cutoff is not yet restrictive. For $N\gg n_L$, $\eta\sim N^{-2}$ instead. 
This behavior is mirrored in the charging power scaling. 
Increasing $n_L$ activates more contributions so that the corresponding power-scaling exponent $\ell(n_L)$ increases from the extensive regime $\ell\simeq 0$, toward the superextensive limit $\ell\simeq 1$, as shown in the inset of~\figlabel{fig:participation}b. 
So, the participation ratio provides a quantitative diagnostic linking the commutator structure of the charger Hamiltonian to the charging power scaling.%

\begin{figure*}[!tb]
    \centering
    \includegraphics[width=\linewidth]{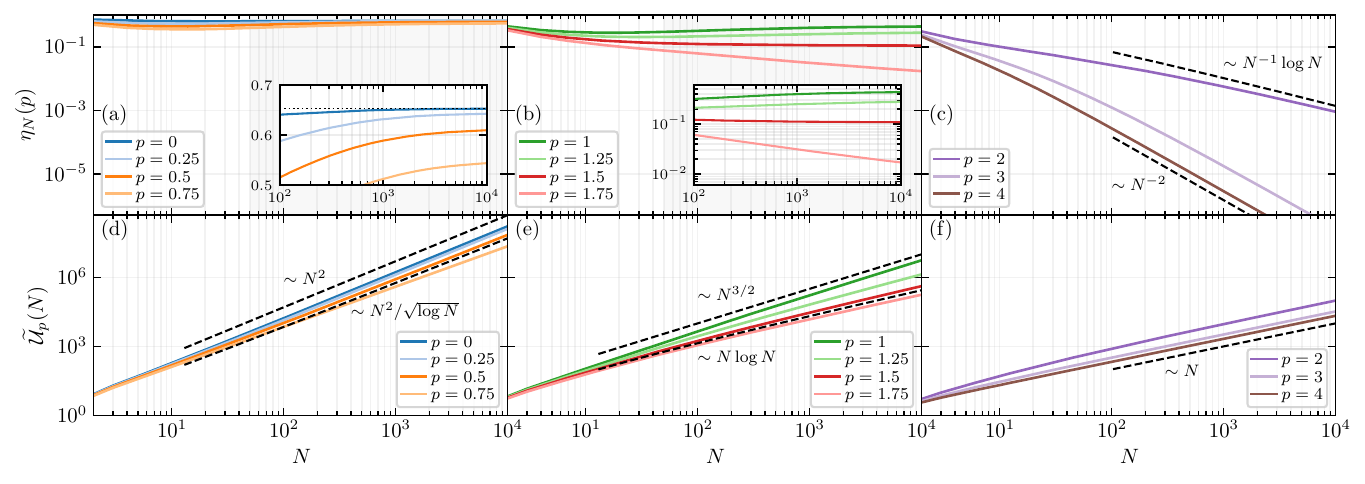}\vspace{-2mm}%
    \caption{\textbf{Scaling under power-law locality-decaying couplings.} Participation ratios $\eta_N(p)$ (\textbf{a}-\textbf{c}) and charging power upper bounds $\widetilde{\mathcal{U}}_p(N)$
    (\textbf{d}-\textbf{f}) for $N\in[2,10000]$ and $p\in[0,4]$. Zoomed insets in \textbf{a} and \textbf{b} magnify the large-$N$ behavior with $\eta_{\rm crit}$ shown as a dotted line in \textbf{a}. All results are averaged over a hundred disorder realizations.%
    }\label{fig:eta_plaw}
\end{figure*}

Cumulative locality can also be accounted in terms of locality-dependent couplings rather than uniform ones. Given the experimental challenge of engineering $k$-local operators, we now explore the same setup but considering power-law locality-decaying couplings in $H_\text{c}(t)$ by performing the map $A_{kl},B_{kl}\mapsto A_{kl}/L^p, B_{kl}/L^p$ with $L=l-k+1$ and $p\ge 0$ the locality-decay exponent. Note that for $p=0$ we recover the previous fully global $n_L=N$ case while for $p\to\infty$ only the local part ($L=1$) survives. Consequently, depending on the $p$ value, a superextensive scaling might be expected or not. Using the same methodology as in~\eqlabel{eq:final_bound} and defining now $\norm{[H_\text{b}, H_\text{c}]}  \le \widetilde{\mathcal{U}}_p(N)$, the different asymptotic regimes are (\sm)
\begin{equation}
\label{eq:U_p}
    \widetilde{\mathcal{U}}_p(N) \sim \begin{cases}
        N^2 & \text{if }0\le p< 1/2 \\
        N^2/\sqrt{\log N} & \text{if }p=1/2 \\
        N^{5/2-p} & \text{if }1/2 < p < 3/2 \\
        N\log N & \text{if }p=3/2 \\
        N & \text{if }p>3/2
    \end{cases}.
\end{equation}
Consequently, the bound is superextensive for $p\le 3/2$, which marginally scales as $N\log N$ for $p=3/2$, and extensively for $p>3/2$. 
The logarithmic corrections correspond to the large-$N$ expansions of the Euler-Maclaurin series (Methods).  
The participation ratio instead exhibits a distinct threshold at $p=2$, scaling as a constant for $p<2$, $N^{-1}\log N$ at $p=2$ and $N^{-2}$ for $p>2$ (\sm). 
Thus, $\eta_N(p)$ serves as a diagnostic of the charging scaling. 
Figure~\ref{fig:eta_plaw} compares both $\widetilde{\mathcal{U}}_p(N)$ bound and $\eta_N(p)$ participation ratios and confirms their predicted asymptotic behaviours.%

Having presented the scaling analyses, we note that our findings are not restricted to free-fermionic Hamiltonians. We illustrate this by considering the Sachdev-Ye-Kitaev (SYK) model and a global charger as examples. We study the scaling of $\norm{[H_\text{b}, H_\text{c}]}$ and $P_*$ for the SYK model [\figlabel{fig:syk_global}a]%
\begin{equation}\label{eq:syk}
    H^{(q)}_\text{c} = i^{q/2} \sum_{i_1 < \dots < i_q} J_{i_1\dots i_q} \chi_{i_1} \dots \chi_{i_q},
\end{equation} 
with $\{\chi_i,\chi_j\}=2\delta_{ij}$, $\chi^2_i=1$ and $J_{i_1 \dots i_q}$ zero-mean Gaussian variables with variance $\sigma^2=N/\binom{2N}{q}$ (Methods), and for the global charger $H_\text{c} = \sum_{s\in\{x,y,z\}} J_s \bigotimes_{i=1}^N \sigma^s_i$ consisting of three $N$-body Pauli strings [\figlabel{fig:syk_global}b], with $J_s$ zero-mean Gaussian variables with $\sigma^2 = N/3$. 
As renormalization imposes $J_s\sim N^{1/2}$, we also study a size-independent variance $\sigma^2=1/3$. We take $H_\text{b}^{(x,z)} = \sum_{i=1}^N \sigma^{(x,z)}_i$ for SYK and $H_\text{b} = \sum_{i=1}^N h_i$ with normally-distributed coefficients for the global charger.%

Analytically proven for $q=2$ in ref.~\cite{divi2025syk}, an $\ell\simeq 1/2$ charging advantage is also expected for $q=4$ in SYK when $H_\text{b}$ contains $\sigma^{(x,y)}$ cells, although extensive scaling is displayed when only $\sigma^z$ terms are considered.
The spin-$1/2$ representation of $\chi_i \chi_j \chi_k \chi_l$ contains $\sigma^-_i S_{ij} \sigma^-_j \sigma^+_k S_{kl} \sigma^+_l$ terms, so the battery-charger commutator excludes noncommuting string-mediated contributions, as in the quadratic subclass discussed above. 
This is consistent with a bosonic SYK-like charger~\cite{rossini2020quantum}, where string operators are absent, yielding at most four-local operators. 
Regarding the global charger, the $\ell\simeq 1/2$ advantage vanishes once the increasing $N$-dependence of its couplings is offset. 
Thus, sustaining the advantage requires the $N$-body couplings to grow with system size, posing a significant challenge for scalable implementations. 
Setting $J_s\sim\mathcal{O}[1]$ instead makes the system behave effectively as $n_L\ll N$, displaying a classically attainable performance, in contrast to the $n_L\propto N$ case despite their coupling strengths decaying as $N^{-1/2}$. This shows that
global $N$-body operators alone do not guarantee a scalable advantage.%

By means of operator norms, we derived structured upper bounds on the instantaneous power of free-fermionic QBs. Beyond global interactions, we prove that a growing set of noncommuting operator families is required for superextensive scaling, although engineering such interactions may be experimentally demanding. 
This finding persists for locality-decaying couplings, for which sufficiently slow decays retain a superextensive upper bound. Importantly, this conclusion extends beyond quadratic free-fermionic chargers, where the charging advantage likewise depends on the proliferation of noncommuting many-body contributions. 
Our findings indicate that achieving a scalable charging advantage through the Hamiltonian structure requires increasingly nonlocal operator structures, which may pose limitations for practical implementations. 
\begin{figure}[!tb]
    \centering
    \includegraphics[width=\linewidth]{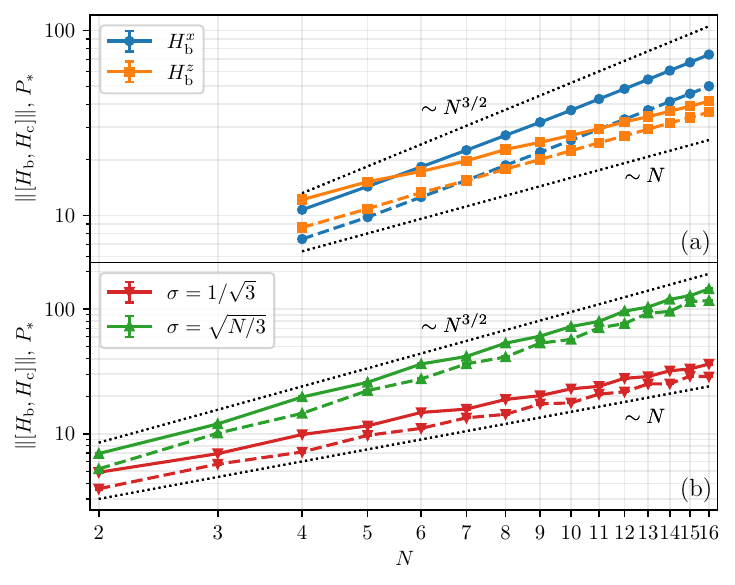}\vspace{-2mm}%
    \caption{
    \textbf{Performance of SYK and global chargers.} Exact operator norms $\norm{[H_\text{b}, H_\text{c}]}$ and maximum charging powers $P_*$ (solid and dashed lines, respectively) 
    are computed for \textbf{a,} the $q=4$ SYK model using the batteries $H_\text{b}^x$ (blue) and $H_\text{b}^z$ (orange); and \textbf{b,} the global charger with $\sigma^2=N/3$ (blue) and $\sigma^2=1/3$ (orange). 
    While SYK results are averaged over $128$ disorder realizations, global charger results are averaged over $5\cdot10^5$ due to the increasing variance with system size.}\label{fig:syk_global}%
\end{figure}%

\emph{Acknowledgements.}---We thank L. Céleri for fruitful discussions. This work is supported by the Spanish Ministry of Science, Innovation, and Universities under Grants PID2024-157842OA-I00 and PID2021-126273NB-I00 funded by MCIN/AEI/10.13039/501100011033 and by ``ERDF A way of making Europe'' and ``ERDF Invest in your Future'', Spanish national project in the field of Artificial Intelligence (AIA2025-163435-C44), and the Severo Ochoa Centres of Excellence program through Grant CEX2024-001445-S.%

\emph{Data and code availability.}---The data that support the findings of this article are openly available~\cite{repository}.

\bibliography{bibfile}

\onecolumngrid
\section*{Methods}
\twocolumngrid

\emph{Derivation of~\eqlabel{eq:bounds}.}---First, we assume that the system evolves unitarily as $\dot{\rho}(t) = -i[H_\text{c}(t),\rho(t)]$. The instantaneous charging power can be computed as
\begin{equation}
\begin{split}
    P(t) &= \trace{\dot{\rho}(t)H_\text{b}} = -i \trace{[H_\text{c}(t),\rho(t)]H_\text{b}} \\
    &= -i\trace{\rho(t)[H_\text{b}, H_\text{c}(t)]},
\end{split}    
\end{equation}
where we use the cyclic property of the trace. From this result, we can use the Hölder inequality~\cite{bhatia1997symmetric}, where%
\begin{equation}
    |\trace{XY}| \le \norm{X}_p \norm{Y}_q \qquad \text{with }\frac{1}{p}+\frac{1}{q}=1,
\end{equation}
being $\norm{\cdot}_r$ the Schatten $r$-norm. By taking $X = \rho(t)$ and $p=1$ ($\norm{\rho(t)}_1 = \trace{\rho(t)} = 1$) and $Y = [H_\text{b}, H_\text{c}(t)]$ and $q\to\infty$, we obtain the leftmost inequality in~\eqlabel{eq:bounds}.%

\emph{Quadratic Hamiltonian renormalization: locality cutoff and coupling strength dependence.}---In order to have a well-defined thermodynamic limit in our analyses, we renormalize the charger Hamiltonian in~\eqlabel{eq:quadratic} to ensure that energies scale linearly with system size. Defining $n_\text{d}$ and $n_\text{od}$ as the number of nonzero diagonal and off-diagonal terms of the $A$ and $B$ matrices, respectively, we have that
\begin{equation}
\begin{split}
    \frac{1}{2^N} \left\langle \trace{H^2_\text{c}(t)} \right\rangle &= \frac{1}{2} \sum_{k=1}^N \left\langle A^2_{kk}(t) \right\rangle \\ &+ \frac{1}{2} \sum_{k< l} \left( \left\langle |A_{kl}(t)|^2  \right\rangle + \left\langle |B_{kl}(t)|^2 \right\rangle \right) \\
    &= \frac{\sigma^2}{2}(n_\text{d} + 2n_\text{od}),
\end{split}
\end{equation}
where $\braket{\cdot}$ denotes the ensemble average, taking zero-mean Gaussian-distributed couplings with variance $\sigma^2$. We can renormalize $H_\text{c}(t)$ by taking $\sigma^2 = 2N/(n_\text{d} + 2n_\text{od})$. In our studies, we use the locality cutoff $n_L$ to control the maximal locality of the charger Hamiltonian terms, which yields $n_\text{d} = N$ and $n_\text{od} = (n_L-1)(2N-n_L)/2$, returning $n_\text{od} = \binom{N}{2}$ for the fully global case ($n_L=N$) and $n_\text{od} = 0$ for the local one ($n_L=1$). 
If we take $n_L \propto N^\nu$ ($0\le \nu\le 1$) instead, we have in the large-$N$ limit that
\begin{equation}
\begin{split}
    \sigma_L^2 &= \frac{2N}{N + (n_L-1)(2N-n_L)} \\ 
    &\approx \frac{1}{n_L}\left[ 1 + \frac{1}{2}\left( \frac{1}{n_L} - \frac{n_L-1}{N} \right) + \mathcal{O}[n_L^{-2}]  \right].
\end{split}
\end{equation}
Substituting $n_L=\kappa N^\nu$, we obtain
\begin{equation}
    \sigma_L^2 \approx \frac{N^{-\nu}}{\kappa} + \frac{N^{-2\nu}}{2\kappa^2} + \frac{N^{-1}}{2} - \frac{N^{-1-\nu}}{2\kappa} + \dots,
\end{equation}
which implies that in the large-$N$ limit $\sigma_L^2\sim \kappa^{-1}N^{-\nu}$, converging to the aforementioned cases when $\nu=0,1$. Apart from the variance, the choice of $\nu$ has an impact on the bound derived in~\theolabel{th:result}. Let $\zeta_n(s)=\sum_{m=1}^n m^{-s}$ be the truncated Riemann zeta function. Using the Euler-Maclaurin expansion, we have for $n$ large that%
\begin{equation}\label{eq:euler_maclaurin}
    \zeta_n(s)  
    = \begin{dcases}
        \frac{n^{1-s}}{1-s} + \zeta(s) + \frac{1}{2n^s} - \frac{s}{12n^{1+s}} + \mathcal{O}\!\left[ \frac{1}{n^{3+s}} \right] & s\neq 1 \\
        \log n + \gamma + \frac{1}{2n} - \frac{1}{12n^2} + \mathcal{O}\!\left[\frac{1}{n^3}\right] & s=1
    \end{dcases},
\end{equation}
with $\zeta(s)=\lim_{n\to\infty}\zeta_n(s)$ the Riemann zeta function and $\gamma\approx 0.5772$ the Euler-Mascheroni constant. % 
In our case, $\sum_{k=1}^{n_L-2}k^{1/2}\sim 2n_L^{3/2}/3$ and $\sum_{k=1}^{n_L-2}k^{3/2}\sim 2n_L^{5/2}/5$, thus
\begin{equation}
    (N-1)\sum_{k=1}^{n_L-2}k^{1/2} - \sum_{k=1}^{n_L-2}k^{3/2} \sim \frac{2}{3}Nn_L^{3/2}.
\end{equation}
Since the rest of contributions in~\eqlabel{eq:final} are subleading in $n_L$ and $N$, its leading term finally reads as
\begin{equation}
    \sqrt{2}\sigma_L \!\left[ 4\Lambda_2\frac{2}{3}Nn_L^{3/2} \right] \approx \frac{8\sqrt{2}}{3}\Lambda_2 \kappa N^{1+\nu} \sim\mathcal{O}[N^{1+\nu}].
\end{equation}

In the limit $n_L\ll N$, we can show that the upper bound presented in~\theolabel{th:result} scales extensively. In this limit, we expand $\sigma_L$ to first order in $n_L/N$, so $\norm{[H_\text{b}, H_\text{c}]} \le P_LN + Q_L$ with%
\begin{widetext}\vspace{-7mm}
\begin{align}
    P_L &= 4b_\perp(n_L-2) + \frac{2}{\sqrt{2n_L-1}}\left[ \Lambda_1 + 2(\Lambda_3+\Lambda_4)\sqrt{n_L-1} \right], \\
    Q_L &= -4b_\perp(n_L-2) + \frac{4(\Lambda_3+\Lambda_4)}{\sqrt{2n_L-1}}\left( \sum_{k=1}^{n_L-1}k^{1/2} - n_L\sqrt{n_L-1}\right) + \frac{n_L(n_L-1)}{(2n_L-1)^{3/2}}\left[ \Lambda_1 + 2(\Lambda_3 + \Lambda_4)\sqrt{n_L-1} \right],
\end{align}\vspace{-4mm}
\end{widetext}%
This result further shows that a locality hierarchy growing with system size is required for a superextensive bound. while linear growth $n_L\propto N$ is required to reach the quadratic limit.

Now, we instead impose a power-law decaying dependence of the coupling strengths on the locality $L=l-k+1$ of their associated operators. $\sigma^2(L) = \sigma^2/L^{2{p}} $
\begin{equation}
\begin{split}
    \frac{1}{2^N} \braket{ \trace{H^2_\text{c}(t)} } &= \frac{\sigma^2}{2}\left( N + 2\sum_{L=2}^N \frac{N+1-L}{L^{2{p}}} \right) \\
    &\implies \sigma^2 = \frac{2N}{N + 2\sum_{L=2}^N \frac{N+1-L}{L^{2{p}}}},
\end{split}
\end{equation}
which can be rewritten in terms of the truncated Riemann zeta function as
\begin{equation}\label{eq:var_power_law}
    \sigma^2 = \frac{2N}{2(N+1)\zeta_N(2{p}) - 2\zeta_N(2{p}-1)-N}.
\end{equation}

Using~\eqlabel{eq:euler_maclaurin} to expand $\zeta_N(s)$ in the large-$N$ limit, we obtain for $0\le {p}< 1/2$ that $\sigma^2\approx 2(1-p)(1-2p)N^{2p-1}$. 
While for ${p}=0$ we get $\sigma^2=2/N$, as obtained before, for ${p}=1/2$ the variance behaves as $\sigma^2 = (\log N + \gamma - 3/2)^{-1}$. For ${p}> 1/2$ and using~\eqlabel{eq:var_power_law}, we have at the large-$N$ limit that $\sigma^2 = 2/[2\zeta(2{p})-1]$, which converges to $\sigma^2=2$ when ${p}\to\infty$, as expected.

\emph{Hill diversity index as effective participation number.}---To estimate effectively the how many terms of the charger Hamiltonian contribute to the local energy injection of each battery cell, we first note that only those noncommuting terms in $[h_k, H_\text{c}(t)] = \sum_{s\in S^{(k)}} c^{(k)}_s P_s$ contribute to the charging. Moreover, we make use of the Hill diversity index the relative strength among them. Obtained by exponentiating the Rényi entropy~\cite{renyi1961measures}, it is defined as
\begin{equation}
    N^{(k)}_a = \Bigg[ \sum_{s\in S^{(k)}} \big( p^{(k)}_s \big)^a \Bigg]^{1/(1-a)},    
\end{equation}
with $p^{(k)}_s$ the normalized weight of the string $s\in S^{(k)}$. In particular, $N^{(k)}_a$ decreases monotonically from $N^{(k)}_{-\infty} = 1/\min_{s\in S^{(k)}} p^{(k)}_s$ (rarest string) to $N^{(k)}_{\infty} = 1/\max_{s\in S^{(k)}} p^{(k)}_s$ (dominant), so varying $a$ merits further exploration. In the $a\to 1$ limit, it reduces to the exponential of the Shannon entropy, $N^{(k)}_1 = \exp\!\big[-\sum_{s\in S^{(k)}} p^{(k)}_s \log p^{(k)}_s \big]$, used in our study. As an example, for $(p_1,p_2)=(0.999,0.001)$ we obtain that $\exp(-\sum_{k=1}^2 p_k\log p_k)\approx 1.008$, while for $(p_1,p_2)=(0.5,0.5)$, $\exp(-\sum_{k=1}^2 p_k\log p_k) = 2$.

\emph{Renormalization of the Sachdev-Ye-Kitaev model.}---We first take as Majorana conventions that $\{\chi_i,\chi_j\}=2\delta_{ij}$ and $\chi^2_i=1$, leading to $\chi_{2i-1} = c^\dagger_i + c_i = -\sigma^x_i\prod_{j<i}\sigma^z_j$ and $\chi_{2i-1} = i(c^\dagger_i - c_i) = -\sigma^y_i\prod_{j<i}\sigma^z_j$. Following the same steps as for the quadratic Hamiltonian, we renormalize the Sachdev-Ye-Kitaev (SYK) model [\eqlabel{eq:syk}] with $2N$ Majorana fermions as%
\begin{equation}
\begin{split}
    \frac{1}{2^N}\braket{\trace{(H^{(q)}_\text{c})^2}} &= \sum_{i_1 < \dots < i_q} \braket{J_{i_1 \dots i_q}^2} = \sigma^2\binom{2N}{q} \\
    &\implies \sigma^2 = \frac{N}{\binom{2N}{q}},
\end{split}
\end{equation}
where the $J_{i_1 \dots i_q}$ coefficients are drawn from a zero-mean $\sigma^2$-variance Gaussian distribution.

The charging performance of $H_\text{b}^{(x,z)}$ can be understood from a Majorana picture $H_\text{b}^x = -\sum_{j=1}^N (-i)^{j-i}\prod_{k=1}^{2j-1}\chi_k$ and $H_\text{b}^z = -i\sum_{j=1}^N \chi_{2j-1}\chi_{2j}$. Note that  although~\eqlabel{eq:syk} is conformed by $q$-local terms, the increasing locality with $N$ is encapsulated in $H_\text{b}^x$, translating into a charging advantage absent in the two-local $H_\text{b}^z$.

As a side note, the charge-zero component of the $q=4$ SYK model is related to the complex SYK $H_\text{cSYK} = \sum_{i,j,k,l=1}^N \mathcal{J}_{ijkl} c^\dagger_i c^\dagger_j c_k c_l$, with $\mathcal{J}_{ijkl}$ zero-mean complex Gaussian variables with $\sigma^2\propto N^{-3}$ satisfying $\mathcal{J}_{ijkl} = -\mathcal{J}_{jikl} = -\mathcal{J}_{ijlk} = \mathcal{J}_{klij}^*$. Let $Q=\sum_{i=1}^N c^\dagger_i c_i$ be the charge operator, with $[Q,H_\text{cSYK}]=0$. 
Decomposing each Majorana as $\chi_a = \chi_a^+ + \chi_a^-$ with $[Q,\chi_a^\pm] = \pm\chi_a^\pm$, the SYK model can be written in terms of charges as%
\begin{equation}
    H_\text{SYK} = \sum_{\Delta=0,\pm2,\pm4} H_{\Delta} = H_0 + \sum_{\Delta=2,4} \big(H_\Delta + H^\dagger_\Delta\big),
\end{equation}
with $[Q, H_\Delta] = \Delta H_\Delta$ and $H_0$ given by
\begin{equation}
    H_0 = \sum_{i,j,k,l=1}^N \mathcal{K}_{ijkl} c^\dagger_i c^\dagger_j c_k c_l + \sum_{i,j=1}^N \mathcal{L}_{ij} c^\dagger_i c_j + \text{const.}
\end{equation}
whose quartic part has the same U$(1)$-conserving structure as $H_\text{cSYK}$. Hence cSYK may be viewed as the quartic U$(1)$-zero-charge component of SYK, known to display a superextensive scaling when $H_\text{b}$ contains $\sigma^{(x,y)}_i$ cells.%

\end{document}

\title{\mbox{{\sm}: ``Cumulative locality governs collective quantum charging''}}

\author{Sebastián V. Romero\orcidlink{0000-0002-4675-4452}}
% \email{sebastian.v.romero@csic.es}
\affiliation{Quantum Advanced Research Center (QuARC), CSIC, 28049 Madrid, Spain}
\affiliation{Instituto de Ciencia de Materiales de Madrid (ICMM), CSIC, 28049 Madrid, Spain}
\affiliation{\mbox{Departamento de Física Teórica de la Materia Condensada, Universidad Autónoma de Madrid, 28049 Madrid, Spain}}

\author{Xi Chen\orcidlink{0000-0003-4221-4288}}
% \email{xi.chen@csic.es}
\affiliation{Quantum Advanced Research Center (QuARC), CSIC, 28049 Madrid, Spain}
\affiliation{Instituto de Ciencia de Materiales de Madrid (ICMM), CSIC, 28049 Madrid, Spain}

\author{Yue Ban\orcidlink{0000-0003-1764-4470}}
% \email{yue.ban@csic.es}
\affiliation{Quantum Advanced Research Center (QuARC), CSIC, 28049 Madrid, Spain}
\affiliation{Instituto de Ciencia de Materiales de Madrid (ICMM), CSIC, 28049 Madrid, Spain}
\date{\today}

\maketitle\vspace{-8mm}

\tableofcontents

\renewcommand{\bibnumfmt}[1]{[S#1]}
\renewcommand{\citenumfont}[1]{S#1}

\let\oldthetable\thetable
\renewcommand{\thetable}{S\oldthetable}

\let\oldthefigure\thefigure
\renewcommand{\thefigure}{S\oldthefigure}

\let\oldtheequation\theequation
\renewcommand{\theequation}{S\oldtheequation}

\let\oldthesection\thesection
\renewcommand{\thesection}{S\oldthesection}

\let\oldthesubsection\thesubsection
\renewcommand{\thesubsection}{S\oldthesubsection}

\section{Pauli decomposition of the charger-battery commutator}

Throughout our derivation, we consider as battery Hamiltonian a general static local Hamiltonian $H_\text{b}=\sum_{i=1}^N h_i$, with $h_i = \alpha_i I_i + \beta_i \sigma^x_i + \gamma_i \sigma^y_i + \delta_i \sigma^z_i$, and a general quadratic form as charger Hamiltonian~\cite{lieb1961two}%
\begin{equation}\label{eq:quadratic}
    H_\text{c}(t) = \sum_{k,l=1}^N \left[ c^\dagger_k A_{kl}(t) c_l + \frac{1}{2}\left( c^\dagger_k B_{kl}(t) c^\dagger_l + \hc \right) \right],
\end{equation}
with $c^\dagger_i$ ($c_i$) a spinless creation (annihilation) fermion at the $i$th site, following the anticommutation rules $\{c_i,c^\dagger_j\}=\delta_{ij}$ and $\{c_i,c_j\}=0$. To preserve Hermiticity, $A$ and $B$ matrices in~\eqlabel{eq:quadratic} have to be Hermitian and skew-symmetric, respectively. Using as Jordan-Wigner transformations
\begin{equation}\label{eq:jwt}
    \sigma^z_i = 1-2c^\dagger_ic_i, \qquad \sigma^+_i = -c_i\prod_{j<i}(1-2c^\dagger_jc_j), \qquad \sigma^-_i = -c^\dagger_i\prod_{j<i}(1-2c^\dagger_jc_j),
\end{equation}
and defining $S_{kl}=\prod_{j=k+1}^{l-1}\sigma^z_j$ as the string operator, we can rewrite the charger Hamiltonian as
\begin{equation}\label{eq:quadratic_pauli}
\begin{split}
    H_\text{c}(t) &= \frac{1}{2}\sum_{k=1}^N A_{kk}(t)(1-\sigma^z_k) +\frac{1}{2}\sum_{k<l} \Big[ (\R{A_{kl}(t)} + \R{B_{kl}(t)})\sigma^x_k S_{kl} \sigma^x_l + (\R{A_{kl}(t)} - \R{B_{kl}(t)})\sigma^y_k S_{kl} \sigma^y_l  \\
    &\qquad\qquad -(\I{A_{kl}(t)} - \I{B_{kl}(t)})\sigma^x_k S_{kl} \sigma^y_l + (\I{A_{kl}(t)} + \I{B_{kl}(t)})\sigma^y_k S_{kl} \sigma^x_l\Big].
\end{split}
\end{equation}

As shown in the main text, the instantaneous power $P(t)$ and battery and charger Hamiltonians are related by the commutator bound $|P(t)|\le \norm{[H_\text{b}, H_\text{c}(t)]}$, which can be directly computed using Pauli commutation identities. 
Expanding the commutator terms into the off-diagonal and diagonal battery Hamiltonian summands as $[H_\text{b}, H_\text{c}(t)] = C_1 + C_2 + C_3 + C_4$, we obtain for the off-diagonal ones $\sum_{i=1}^N [\beta_i\sigma^x_i + \gamma_i\sigma^y_i, H_\text{c}(t)] = C_1 + C_2 + C_3$ with
\begin{align}\label{eq:c1}
    C_1 &= i\sum_{k=1}^N \left( \kappa_{1,1}^{(k)}\sigma^x_k + \kappa_{1,2}^{(k)}\sigma^y_k \right) \qquad(\text{with } \kappa_{1,1}^{(k)} = -A_{kk}\gamma_k\text{ and } \kappa_{1,2}^{(k)} = A_{kk}\beta_k), \\ \label{eq:c2}
    \begin{split}
    C_2 &= i\sum_{k<m}\sum_{l=k+1}^{m-1} \!\!\!\Big( \kappa_{2,1}^{(klm)}\sigma^x_k S_{kl} \sigma^x_l S_{lm} \sigma^x_m + \kappa_{2,2}^{(klm)}\sigma^x_k S_{kl} \sigma^y_l S_{lm} \sigma^x_m + \kappa_{2,3}^{(klm)}\sigma^y_k S_{kl} \sigma^y_l S_{lm} \sigma^y_m + \kappa_{2,4}^{(klm)}\sigma^y_k S_{kl} \sigma^x_l S_{lm} \sigma^y_m \\
    &\qquad\qquad + \kappa_{2,5}^{(klm)}\sigma^x_k S_{kl} \sigma^x_l S_{lm} \sigma^y_m + \kappa_{2,6}^{(klm)}\sigma^x_k S_{kl} \sigma^y_l S_{lm} \sigma^y_m + \kappa_{2,7}^{(klm)}\sigma^y_k S_{kl} \sigma^x_l S_{lm} \sigma^x_m + \kappa_{2,8}^{(klm)}\sigma^y_k S_{kl} \sigma^y_l S_{lm} \sigma^x_m \Big)
    \end{split} \\ \label{eq:c3}
    C_3 &= i\sum_{k<l} \left( \kappa_{3,1}^{(kl)}S_{k-1,l}\sigma^x_l + \kappa_{3,2}^{(kl)}S_{k-1,l}\sigma^y_l + \kappa_{3,3}^{(kl)}\sigma^x_k S_{k,l+1} + \kappa_{3,4}^{(kl)}\sigma^y_k S_{k,l+1} \right).
\end{align}
The coefficients for the $C_2$ family are given by
\begin{equation}\label{eq:coeffs_c2}
\begin{aligned}
    \kappa_{2,1}^{(klm)} &= \gamma_l\R{A_{km} + B_{km}}, & \kappa_{2,2}^{(klm)} &= -\beta_l\R{A_{km} + B_{km}}, & \kappa_{2,3}^{(klm)} &= -\beta_l\R{A_{km} - B_{km}}, & \kappa_{2,4}^{(klm)} &= \gamma_l\R{A_{km} - B_{km}}, \\
    \kappa_{2,5}^{(klm)} &= -\gamma_l\I{A_{km} - B_{km}}, & \kappa_{2,6}^{(klm)} &= \beta_l\I{A_{km} - B_{km}}, & \kappa_{2,7}^{(klm)} &= \gamma_l\I{A_{km} + B_{km}}, & \kappa_{2,8}^{(klm)} &= -\beta_l\I{A_{km} + B_{km}},
\end{aligned}
\end{equation}
while the coefficients for the $C_3$ family are given by
\begin{equation}\label{eq:coeffs_c3}
\begin{aligned}
    \kappa_{3,1}^{(kl)} &= \beta_k\I{A_{kl}+B_{kl}} -\gamma_k\R{A_{kl}+B_{kl}}, &\quad \kappa_{3,2}^{(kl)} &= \beta_k\R{A_{kl}-B_{kl}} + \gamma_k\I{A_{kl}-B_{kl}}, \\
    \kappa_{3,3}^{(kl)} &= -\beta_l\I{A_{kl}-B_{kl}} -\gamma_l\R{A_{kl}+B_{kl}}, &\quad \kappa_{3,4}^{(kl)} &= \beta_l\R{A_{kl}-B_{kl}} - \gamma_l\I{A_{kl}+B_{kl}}.
\end{aligned}
\end{equation}
We group terms for convenience, seeing that for each $(k,l,m)$ set of indices, $C_1$, $C_2$ and $C_3$ matrices have Pauli strings sharing off-diagonal Pauli matrices at the same spin indices, which will be helpful for computing their norms later.

Proceeding similarly with the diagonal part of $H_\text{b}$, where $\sum_{i=1}^N [\delta_i\sigma^z_i, H_\text{c}(t)] = C_4$, we obtain that
\begin{equation}\label{eq:c4}
    C_4 = i\sum_{k<l} \left( \kappa_{4,1}^{(kl)}\str{x}{x} + \kappa_{4,2}^{(kl)}\str{x}{y} + \kappa_{4,3}^{(kl)}\str{y}{x} + \kappa_{4,4}^{(kl)}\str{y}{y} \right),
\end{equation}
with coefficients
\begin{equation}\label{eq:coeffs_c4}
\begin{aligned}
    \kappa_{4,1}^{(kl)} &= -\delta_k\I{A_{kl}+B_{kl}} + \delta_l\I{A_{kl}-B_{kl}}, &\quad \kappa_{4,2}^{(kl)} &= -\delta_k\R{A_{kl}-B_{kl}} + \delta_l\R{A_{kl}+B_{kl}}, \\
    \kappa_{4,3}^{(kl)} &= \delta_k\R{A_{kl}+B_{kl}} - \delta_l\R{A_{kl}-B_{kl}}, &\quad \kappa_{4,4}^{(kl)} &= -\delta_k\I{A_{kl}-B_{kl}} + \delta_l\I{A_{kl}+B_{kl}}.
\end{aligned}
\end{equation}

Notice that the identity-related terms of the battery Hamiltonian are not present, since they only introduce a constant shift per spin, thus they trivially commute. As for the $C_1$, $C_2$ and $C_3$ matrices, see also that for each pair of indices $(k,l)$, the resulting Pauli strings share off-diagonal Pauli matrices at the same spin indices. Therefore, we can rewrite each $C_i$ matrix as a sum of those terms whose off-diagonal matrices Pauli matrices are located at the same spin indices in equations~\eqref{eq:c1}-\eqref{eq:c4}, namely $C_1 = \sum_{k=1}^N C^{(k)}_1$, $C_2 = \sum_{k<m} \sum_{l=k+1}^{m-1} C^{(klm)}_2$, and $C_{\{3,4\}} = \sum_{k<l} C^{(kl)}_{\{3,4\}}$.%

\section{Calculation of the charging power upper bound}\label{sec:upper_bound}

In order to derive an upper on the charging power $P(t)$, we first use that $|P(t)| \le \norm{[H_\text{b}, H_\text{c}(t)]} \le \sum_{j=1}^4 \norm{C_j}$. Prior to estimating these norms and in order to have a well-defined thermodynamic limit, we renormalize the charger Hamiltonian in~\eqlabel{eq:quadratic} to ensure that energies scale linearly with system size. Defining $n_\text{d}$ and $n_\text{od}$ as the number of nonzero diagonal and off-diagonal terms of the $A$ and $B$ matrices, respectively, we have that
\begin{equation}\label{eq:renormalization}
    \frac{1}{2^N} \left\langle \trace{H^2_\text{c}(t)} \right\rangle = \frac{1}{2} \sum_{k=1}^N \left\langle A^2_{kk}(t) \right\rangle + \frac{1}{2} \sum_{k< l} \left( \left\langle |A_{kl}(t)|^2  \right\rangle + \left\langle |B_{kl}(t)|^2 \right\rangle \right) = \frac{\sigma^2}{2}(n_\text{d} + 2n_\text{od}),
\end{equation}
where $\braket{\cdot}$ denotes the ensemble average, taking zero-mean Gaussian-distributed couplings with variance $\sigma^2$. We can renormalize $H_\text{c}(t)$ by taking $\sigma^2 = 2N/(n_\text{d} + 2n_\text{od})$. 
If we use the locality cutoff $n_L$ to control the maximal locality of the charger Hamiltonian terms, we get $n_\text{d} = N$ and $n_\text{od} = \sum_{k=2}^{n_L} (N-k+1) = (n_L-1)(2N-n_L)/2$, returning $n_\text{od} = \binom{N}{2}$ for the fully global case ($n_L=N$) and $n_\text{od} = N-1$ for the tight-binding one ($n_L=2$). Moreover, decomposing $A_{kl} = \sigma_L \tilde{A}_{kl}$ and $B_{kl} = \sigma_L \tilde{B}_{kl}$ with $\mathbb{E}[|\tilde{A}_{kl}|^2] = \mathbb{E}[|\tilde{B}_{kl}|^2] = 1$, we define $\Lambda_1= \max_{k\nu} |\tilde{\kappa}_{1,\nu}^{(k)}|$ ($\nu=1,2$) and $\Lambda_{(3,4)} = \max_{kl\nu} |\tilde{\kappa}^{(kl)}_{(3,4),\nu}|$ ($\nu=1,2,3,4$) the largest Pauli coefficients in absolute value of the unnormalized $C_1$ and $C_{(3,4)}$ families, respectively.

The operator norm of $C_1$ can be directly computed as $\norm{C_1} = \sum_{k=1}^N \sqrt{(\kappa_{1,1}^{(k)})^2 + (\kappa_{1,2}^{(k)})^2}$. However, in our scaling analyses we make use of the parameter $n_L$, which serves as a locality cutoff to include all terms in $H_\text{c}(t)$ ranging from local to $n_L$-body operators. We can rewrite an upper bound in terms of $n_L$, which can be renormalized using $\sigma^2_L=2N/[N+(n_L-1)(2N-n_L)]$. We then obtain that $\norm{C_1} \le \sqrt{2}\Lambda_1\sigma_L N$. For $n_L=N$, it transforms into $\norm{C_1}\le 2\Lambda_1\sqrt{N}$, since $\sigma^2_N=2/N$.

To compute an upper bound on the $C_2$ norm, we first rewrite $C_2 = \sum_{l=2}^{N-1} [b_l, K_l]$ as sum of commutators with $b_l = \beta_l\sigma^x_l + \gamma_l\sigma^y_l$ and $K_l = \sum_{\substack{k<l<m\\m-k\le n_L-1}} H_{km}$ ($\forall l\in[2,N-1]$ and $n_L\ge 3$), where
\begin{equation}
\begin{split}
    H_{km} &= \frac{1}{2} \Big[ (\R{A_{km}} + \R{B_{km}})\sigma^x_k S_{km} \sigma^x_m + (\R{A_{km}} - \R{B_{km}})\sigma^y_k S_{km} \sigma^y_m  \\
    &\qquad\qquad -(\I{A_{km}} - \I{B_{km}})\sigma^x_k S_{km} \sigma^y_m + (\I{A_{km}} + \I{B_{km}})\sigma^y_k S_{km} \sigma^x_m\Big].
\end{split}
\end{equation}
By means of the triangle inequality and submultiplicativity of the operator norm, we can write
\begin{equation}
    \norm{C_2} \le 2\sum_{l=2}^{N-1} \norm{b_l}\norm{K_l} = 2\sum_{l=2}^{N-1} \sqrt{\beta_l^2 + \gamma_l^2} \norm{K_l}.
\end{equation}
Now we bound each $K_l$ norm in terms of the $A$ and $B$ matrices. For a complex rectangular matrix $M$, we define its nuclear norm as $\norm{M}_* = \sum_{j=1}^r s_j(M)$, with $s_j(M)>0$ the nonzero singular values and $r=\text{rank}(M)$. Consider two disjoint sets of fermionic modes $a_1, \dots, a_p$ and $d_1, \dots, d_s$. Define the Hermitian hopping and pairing operators
\begin{equation}
    T(M) = \sum_{i=1}^p \sum_{j=1}^s \left( M_{ij} a_i^\dagger d_j + M_{ij}^* d_j^\dagger a_i \right), \qquad Q(M) = \sum_{i=1}^p \sum_{j=1}^s \left( M_{ij} a_i^\dagger d_j^\dagger + M_{ij}^* d_j a_i \right).
\end{equation}
We can bound the norm of $K_l$ by means of the norms of the pairing and hopping operators. Starting from the hopping operator, let $M = U D V^\dagger$ a singular value decomposition, with nonzero diagonal entries $s_1, \dots, s_r$ in $D$. The unitary change $a = U\tilde{a}$ and $d = V\tilde{d}$ transforms $T(M) = \sum_{j=1}^r s_jX_j$ with $X_j = \tilde{a}_j^\dagger \tilde{d}_j + \tilde{d}_j^\dagger \tilde{a}_j$ and $\norm{X_j}=1$. Therefore, it follows from the triangle inequality that $\norm{T(M)} \le \sum_{j=1}^r s_j$. 
To prove the reverse inequality, let $\ket{\Omega}$ be the vacuum state and define the normalized state $\ket{\psi_T} = 2^{-r/2} \prod_{j=1}^r \big( \tilde{a}_j^\dagger + \tilde{d}_j^\dagger \big)\ket{\Omega}$. Since $X_j$ commutes with other mode pairs, $X_j\ket{\psi_T} = \ket{\psi_T}$ $\forall j$. Consequently,
\begin{equation}
    T(M)\ket{\psi_T} = \sum_{j=1}^r s_j \ket{\psi_T} \implies \norm{T(M)} \ge \norm{T(M)\ket{\psi_T}} = \sum_{j=1}^r s_j .
\end{equation}
Combining both upper and lower bounds, we finally conclude that $\norm{T(M)} = \sum_{j=1}^r s_j = \norm{M}_*$.

Continuing with the pairing operator, for $Q(M)$ we consider the singular value decomposition $M = U D V^\dagger$ and the unitary mode transformations $a = U\tilde{a}$ and now $d = V^*\tilde{d}$, implying that $Q(M) = \sum_{j=1}^r s_j Y_j$ where $Y_j = \tilde{a}_j^\dagger \tilde{d}_j^\dagger + \tilde{d}_j \tilde{a}_j$ with $\norm{Y_j}=1$. Following a similar procedure with as for the hopping operator, with common normalized eigenstate $\ket{\psi_Q} = 2^{-r/2}\prod_{j=1}^r (1+\tilde{a}_j^\dagger\tilde{d}_j^\dagger)\ket{\Omega}$ with $Y_j\ket{\psi_Q} = \ket{\psi_Q}$ $\forall j$, we have that $\norm{Q(M)} = \sum_{j=1}^r s_j = \norm{M}_*$.%

To apply these fermionic norms to $K_l$, we define the rectangular matrices $A^{(l)}_{km}$ and $B^{(l)}_{km}$ of dimension $(l-1)\times(N-l)$%
\begin{equation}
    A^{(l)}_{km} = \begin{cases}
        A_{km} & \text{if }m-k\le n_L-1 \\ 0 & \text{otherwise}
    \end{cases}, \qquad B^{(l)}_{km} = \begin{cases}
        B_{km} & \text{if }m-k\le n_L-1 \\ 0 & \text{otherwise}
    \end{cases}.
\end{equation}

The hopping terms in $K_l$ are exactly $T(A^{(l)}_{km})$ while the pairing ones are $Q(B^{(l)}_{km})$. Therefore $K_l = T(A^{(l)}_{km}) + Q(B^{(l)}_{km})$, implying that $\norm{K_l} \le \nnorm{T(A^{(l)}_{km})} + \nnorm{Q(B^{(l)}_{km})} = \nnorm{A^{(l)}_{km}}_* + \nnorm{B^{(l)}_{km}}_*$. The nuclear norms can be bounded by the Frobenius norm $\norm{M}_\text{F} = \big( \sum_{i,j} |M_{ij}|^2 \big)^{1/2}$ as $\norm{M}_* \le \sqrt{\text{rank}(M)} \norm{M}_\text{F}$. Let $r_l=\min\{n_L-2, l-1, N-l\}$. Since the rank cannot exceed either the number of nonzero rows or columns, this fact yields that $\nnorm{A^{(l)}_{km}}_*\le \sqrt{r_l} \nnorm{A^{(l)}_{km}}_\text{F}$ and $\nnorm{B^{(l)}_{km}}_*\le \sqrt{r_l} \nnorm{B^{(l)}_{km}}_\text{F}$. 
\begin{equation}
    \norm{C_2} \le 2\sum_{l=2}^{N-1} \sqrt{\beta_l^2 + \gamma_l^2} \left( \nnorm{A^{(l)}_{km}}_* + \nnorm{B^{(l)}_{km}}_* \right) \le 2\sum_{l=2}^{N-1} \sqrt{\beta_l^2 + \gamma_l^2} \sqrt{r_l} \left( \nnorm{A^{(l)}_{km}}_\text{F} + \nnorm{B^{(l)}_{km}}_\text{F} \right).
\end{equation}
Applying Cauchy-Schwarz over the $2(N-2)$ terms gives
\begin{equation}
\begin{split}
    \sum_{l=2}^{N-1} \sqrt{\beta_l^2 + \gamma_l^2} \sqrt{r_l} \left( \nnorm{A^{(l)}_{km}}_\text{F} + \nnorm{B^{(l)}_{km}}_\text{F} \right) &\le \sqrt{2 \sum_{l=2}^{N-1} ( \beta_l^2 + \gamma_l^2 ) r_l} \sqrt{\sum_{l=2}^{N-1} \left( \nnorm{A^{(l)}_{km}}_\text{F}^2 + \nnorm{B^{(l)}_{km}}_\text{F}^2 \right) } \\
    &= \sqrt{2 \sum_{l=2}^{N-1} ( \beta_l^2 + \gamma_l^2 ) r_l} \sqrt{\sum_{\substack{k<m\\2\le m-k\le n_L-1}} (m-k-1)\left( |A_{km}|^2 + |B_{km}|^2 \right) },
\end{split}
\end{equation}
which already provides a bound in terms of the system parameters. Using that $m-k-1\le n_L-2$, $r_l\le n_L-2$ and defining $b^2_\perp = \max_{2\le l\le N-1} (\beta_l^2 + \gamma_l^2)$, we obtain that
\begin{equation}\label{eq:tight_bound_c2}
    \norm{C_2} \le 2\sqrt{2}b_\perp (n_L-2) \sqrt{(N-2)\sum_{\substack{k<m\\2\le m-k\le n_L-1}} \left( |A_{km}|^2 + |B_{km}|^2 \right) }.
\end{equation}
This final expression can be further reduced by using the charger normalization. 
The charger Hamiltonian $H_\text{c} = H_\text{c}^\text{d} + H_\text{c}^\text{od}$ splits into its diagonal and off-diagonal parts as
\begin{equation}
    H_\text{c}^\text{d} = \frac{1}{2}\sum_{k=1}^N A_{kk}(1-\sigma^z_k), \qquad H_\text{c}^\text{od} = \sum_{\substack{k<m\\m-k\le n_L-1}} H_{km}.
\end{equation}

Renormalization and Pauli string orthogonality impose that
\begin{equation}
    \frac{\braket{\trace{(H_\text{c})^2}}}{2^N} = \frac{\braket{\trace{(H_\text{c}^\text{d})^2 + (H_\text{c}^\text{od})^2}}}{2^N} = N \implies \frac{\braket{\trace{(H_\text{c}^\text{od})^2}}}{2^N} = \frac{1}{2} \sum_{\substack{k<m\\m-k\le n_L-1}} \left( \braket{|A_{km}|^2} + \braket{|B_{km}|^2} \right) \le N.
\end{equation}
Substituting back into~\eqlabel{eq:tight_bound_c2}, we finally obtain as upper bound for an ensemble average that
\begin{equation}\label{eq:bound_c2}
    \mathbb{E}[\norm{C_2}] \le 4b_\perp (n_L-2)\sqrt{N(N-2)} \sim \mathcal{O}[Nn_L]
\end{equation}
which, as expected, displays an $N^2$ superextensive scaling for the $n_L=N$ case while for constant $n_L$ it provides an asymptotic extensive scaling.

Regarding the remaining $C_{(3,4)}$ families, we can group terms into pairwise anticommuting subfamilies such that their squares correspond to the identity matrix times the operator norm squared.
\begin{theorem}\label{th:pairwise}
    Let $M=\sum_{j=1}^m a_j P_j$, where $a_j\in\mathbb{R}$ and $P_j$ is a Pauli string, satisfying $P_j^\dagger=P_j$ and $P_j^2=I$, and assume that the family $\{P_j\}_{j=1}^m$ is pairwise anticommuting, thus $P_jP_k+P_kP_j=0$ for all $j\neq k$. Then, $\norm{M}=\sqrt{\sum_{j=1}^m a_j^2}$.
\end{theorem}%
\begin{proof}
Expand the square, which returns
\begin{equation}
    M^2=\Bigg(\sum_{j=1}^m a_jP_j\Bigg)^2 =\sum_{j=1}^m a_j^2P_j^2+\sum_{j\neq k}a_ja_kP_jP_k = \sum_{j=1}^m a_j^2 I + \sum_{j<k}a_ja_k(P_jP_k + P_kP_j) = \sum_{j=1}^m a_j^2 I.
\end{equation}
Because $M$ is Hermitian, its operator norm satisfies $\norm{M}^2=\norm{M^2}=\sum_{j=1}^m a_j^2$. Therefore, $\norm{M}=\sqrt{\sum_{j=1}^m a_j^2}$.
\end{proof}

Proceeding with the $C_3$ family, we can split $C_3$ into two subfamilies of pairwise anticommuting matrices: one setting the leftmost spin index $k$ and another one setting the rightmost index $l$. Concretely,
\begin{equation}\label{eq:c3_pairwise}
    \norm{C_3} = \norm{ \sum_{k=1}^{N-1}\sum_{l=k+1}^N C_3^{(kl)} } = \norm{ \sum_{k=1}^{N-1}\sum_{l=k+1}^{\min(k+n_L-1,N)} C_3^{(kl)} } \le \sum_{k=1}^{N-1} \norm{C_{3,k}^L} + \sum_{l=2}^N \norm{C_{3,l}^R},
\end{equation}
where, using the coefficients presented in~\eqlabel{eq:coeffs_c3}, we define
\begin{equation}
    \norm{C_{3,k}^L}^2 = \sum_{l=k+1}^{\min(k+n_L-1,N)} \left[ (\kappa_{3,1}^{(kl)})^2 + (\kappa_{3,2}^{(kl)})^2 \right], \qquad \norm{C_{3,l}^R}^2 = \sum_{k=\max(1,l-n_L+1)}^{l-1} \left[ (\kappa_{3,3}^{(kl)})^2 + (\kappa_{3,4}^{(kl)})^2 \right].
\end{equation}
This already provides an upper bound on the operator norm of $C_3$ after substituting it back into~\eqlabel{eq:c3_pairwise}. We can reflect the locality cutoff $n_L$ in this bound more explicitly as
\begin{equation}\label{eq:c3_nl_norm}
\begin{split}
    \norm{C_3} &\le \sum_{k=1}^{N-1} \norm{C_{3,k}^L} + \sum_{l=2}^N \norm{C_{3,l}^R} \le 2\sqrt{2}\Lambda_3\sigma_L \left[\sum_{k=1}^{n_L-1}k^{1/2} + (N-n_L)\sqrt{n_L-1} \right].
\end{split}
\end{equation}
For $n_L=N$ with $N\to\infty$, we can expand the two summands in terms of an Euler-Maclaurin series as
\begin{equation}\label{eq:euler_maclaurin}
    \sum_{m=1}^n \frac{1}{m^s} = \begin{dcases}
        \frac{n^{1-s}}{1-s} + \zeta(s) + \frac{1}{2n^s} - \frac{s}{12n^{1+s}} + \mathcal{O}\!\left[ \frac{1}{n^{3+s}} \right] & \text{if }s\neq 1 \\
        \log n + \gamma + \frac{1}{2n} - \frac{1}{12n^2} + \mathcal{O}\!\left[\frac{1}{n^3}\right] & \text{if }s=1
    \end{dcases}
\end{equation}
with $n= N-1$, where
\begin{equation}
    (N-a)^r = N^r \sum_{j=0}^{\lfloor r+3/2\rfloor} \binom{r}{j} \left(-\frac{a}{N}\right)^j + \mathcal{O}[N^{-5/2}], \qquad r\in\left\{\frac52,\frac32,\frac12,-\frac12\right\}
\end{equation}
with $a\in\mathbb{R}$. Hence we finally obtain that
\begin{equation}
    \norm{C_3} \le 4\Lambda_3 N^{-1/2} \sum_{k=1}^{N-1}k^{1/2} = 2\Lambda_3 \left( \frac{4}{3}N - 1 \right) + \mathcal{O}[N^{-1/2}].
\end{equation}
Therefore, the $C_3$-family contribution saturates in this limit to an extensive scaling in the charging power, as expected.%

We conclude by computing the norm upper bound of $C_4$, which can also be split into two subfamilies after setting the leftmost spin index $k$: one with $\sigma^x_k$ as leftmost Pauli matrix endpoint and another one with $\sigma^y_k$. In particular,
\begin{equation}\label{eq:c4_pairwise}
    \norm{C_4} = \norm{ \sum_{k=1}^{N-1}\sum_{l=k+1}^N C_4^{(kl)} } = \norm{ \sum_{k=1}^{N-1}\sum_{l=k+1}^{\min(k+n_L-1,N)} C_4^{(kl)} } \le \sum_{k=1}^{N-1} \left( \norm{C_{4,k}^x} + \norm{C_{4,k}^y} \right),
\end{equation}
where we have used the parameters presented in~\eqlabel{eq:coeffs_c4} and define
\begin{equation}
    \norm{C_{4,k}^x}^2 = \sum_{l=k+1}^{\min(k+n_L-1,N)} \left[ (\kappa_{4,1}^{(kl)})^2 + (\kappa_{4,2}^{(kl)})^2 \right], \qquad \norm{C_{4,k}^y}^2 = \sum_{l=k+1}^{\min(k+n_L-1,N)} \left[ (\kappa_{4,3}^{(kl)})^2 + (\kappa_{4,4}^{(kl)})^2 \right].
\end{equation}
We can write this bound in terms of the locality cutoff $n_L$ as
\begin{equation}
\begin{split}
    \norm{C_4} &\le \sum_{k=1}^{N-1} \left( \norm{C_{4,k}^x} + \norm{C_{4,k}^y} \right) \le 2\sqrt{2}\Lambda_4\sigma_L \left[\sum_{k=1}^{n_L-1}k^{1/2} + (N-n_L)\sqrt{n_L-1} \right],
\end{split}
\end{equation}
whose form is similar to the upper bound obtained for $C_3$ (see~\eqlabel{eq:c3_nl_norm}). Setting $n_L=N$ and taking the thermodynamic limit $N\to\infty$, we can rewrite this bound using~\eqlabel{eq:euler_maclaurin} expansion as
\begin{equation}
\begin{split}
    \norm{C_4} \le 4\Lambda_4 N^{-1/2}\sum_{k=1}^{N-1} k^{1/2} = 2\Lambda_4\left(\frac{4}{3}N - 1 \right) + \mathcal{O}[N^{-1/2}].
\end{split}
\end{equation}

Summing all upper bounds together and using as in the main text that $|P(t)|\le \norm{[H_\text{b}, H_\text{c}(t)]} \le \mathcal{U}_{n_L}(N)$, we get%
\begin{equation}
    \mathcal{U}_{n_L}(N)= 4b_\perp (n_L-2)\sqrt{N(N-2)} + 
    \sqrt{2}\sigma_L \left[ \Lambda_1 N +2(\Lambda_3 + \Lambda_4)\left[ \sum_{k=1}^{n_L-1} k^{1/2} + (N-n_L)\sqrt{n_L-1} \right] \right] 
\end{equation}%
with fully global $n_L=N$ asymptotic limit
\begin{equation}
    \mathcal{U}_{N}(N)=
    4b_\perp N^2 + \left[ \frac{8}{3}(\Lambda_3 + \Lambda_4) -12b_\perp \right] N + 2 \Lambda_1 N^{1/2} + 2\left( 3b_\perp - \Lambda_3 - \Lambda_4 \right) + \mathcal{O}[N^{-1/2}],
\end{equation}
which corresponds to the Theorem 1 presented in the main text. In the limit $n_L\ll N$, this upper bound scales extensively with system size. Since $\sigma^2_L\to 2/(2n_L-1)$ in this limit, $\norm{[H_\text{b}, H_\text{c}]} \le P_LN + Q_L$ with
\begin{align}
    P_L &= 4b_\perp(n_L-2) + \frac{2}{\sqrt{2n_L-1}}\left[ \Lambda_1 + 2(\Lambda_3+\Lambda_4)\sqrt{n_L-1} \right], \\
    Q_L &= -4b_\perp(n_L-2) + \frac{4(\Lambda_3+\Lambda_4)}{\sqrt{2n_L-1}}\left( \sum_{k=1}^{n_L-1}k^{1/2} - n_L\sqrt{n_L-1}\right) + \frac{n_L(n_L-1)}{(2n_L-1)^{3/2}}\left[ \Lambda_1 + 2(\Lambda_3 + \Lambda_4)\sqrt{n_L-1} \right],
\end{align}
which stresses the need of cumulative locality $n_L\propto N$ to reach a scalable superextensive scaling in the charging power.%

\subsection{Quadratic form for a diagonal battery Hamiltonian}

Note that when the battery Hamiltonian $H_\text{b}$ is diagonal, i.e., $\beta_i, \gamma_i = 0$ $\forall i\in[1,N]$, the commutator $[H_\text{b}, H_\text{c}(t)]$ remains quadratic. Namely, using~\eqlabel{eq:jwt}, it can be rewritten as $[H_\text{b}, H_\text{c}(t)] = -2\sum_{i=1}^N\delta_i[c^\dagger_ic_i,H_\text{c}(t)]$. For its computation, the key identities to be used are that $[c^\dagger_i c_i, c^\dagger_k c_l] = (\delta_{ik} - \delta_{il}) c^\dagger_k c_l$, $[c^\dagger_i c_i, c^\dagger_k c^\dagger_l] = (\delta_{ik} + \delta_{il}) c^\dagger_k c^\dagger_l$, and $[c^\dagger_i c_i, c_l c_k] = -(\delta_{ik} + \delta_{il}) c_l c_k$. 
Substituting these identities into the commutator, we obtain 
\begin{equation}
\begin{split}
    [H_\text{b}, H_\text{c}(t)] &= -2\sum_{i=1}^N \delta_i \sum_{k,l=1}^N \left[ A_{kl}(t) [c^\dagger_i c_i, c^\dagger_k c_l] + \frac{1}{2} \left( B_{kl}(t) [c^\dagger_i c_i, c^\dagger_k c^\dagger_l] + B^*_{kl}(t) [c^\dagger_i c_i, c_l c_k] \right) \right] \\
    &= -2\sum_{i=1}^N \delta_i \sum_{k,l=1}^N \left[ A_{kl}(t) (\delta_{ik} - \delta_{il}) c^\dagger_k c_l + \frac{1}{2} \left( B_{kl}(t) (\delta_{ik} + \delta_{il}) c^\dagger_k c^\dagger_l - B^*_{kl}(t) (\delta_{ik} + \delta_{il}) c_l c_k \right) \right] \\
    &= -2 \sum_{k,l=1}^N \left[ A_{kl}(t) (\delta_k - \delta_l) c^\dagger_k c_l + \frac{\delta_k + \delta_l}{2} \left( B_{kl}(t) c^\dagger_k c^\dagger_l - B^*_{kl}(t) c_l c_k \right) \right].
\end{split}
\end{equation}

Defining $\Psi = [c_1,\dots,c_N,c^\dagger_1,\dots,c^\dagger_N]^\text{T}$ and $D= \diag(\delta_1,\dots,\delta_N)$, the commutator can be finally rewritten as
\begin{equation}\label{eq:c}
    [H_\text{b}, H_\text{c}(t)] = \frac{1}{2} \Psi^\dagger \mathcal{C}(t) \Psi \qquad\text{with }\mathcal{C}(t) = \begin{bmatrix} \mathcal{A}(t) & \mathcal{B}(t) \\ \mathcal{B}^*(t) & -\mathcal{A}^\text{T}(t) \end{bmatrix},
\end{equation}
and $\mathcal{A}(t) = 2[A(t),D]$ and $\mathcal{B}(t) = -2\{B(t),D\}$, satisfying that $\mathcal{A}^\dagger(t)=-\mathcal{A}(t)$ and $\mathcal{B}^\text{T}(t)=-\mathcal{B}(t)$. Since the commutator is quadratic in fermions, it exists a Bogoliubov transformation $d=Uc+Vc^\dagger$ that diagonalizes it as $[H_\text{b}, H_\text{c}(t)] = i\sum_{j=1}^N \epsilon_j (d^\dagger_j d_j - \frac{1}{2})$ with $\epsilon_j\ge 0$ and $n_j=d^\dagger_j d_j=0,1$ $\forall j$. Therefore, the largest singular value of the commutator can be computed as $\norm{[H_\text{b}, H_\text{c}(t)]} = \frac{1}{2} \sum_{j=1}^N \epsilon_j = \frac{1}{4} \sum_{j=1}^{2N} |\lambda_j|$, with $\lambda_j=\pm i\epsilon_j$ the eigenvalues of $\mathcal{C}(t)$.%

We can also leverage the quadratic structure of $H_\text{c}$ to compute $P(t)$ and write
\begin{equation}
    H_\text{c} = \frac{1}{2}\Psi^\dagger \mathcal{G}\Psi \qquad\text{with }\mathcal{G}=\begin{bmatrix} A & B \\ -B^* & -A^\text{T} \end{bmatrix}
\end{equation}
and use that the charging power can be computed as $P(t)=\trace{\dot{\rho}(t)H_\text{b}} = -i\trace{\rho(0)e^{-iH_\text{c}t}[H_\text{b},H_\text{c}]e^{iH_\text{c}t}}$. A natural choice of the initial state $\rho(0)$ is the ground state of $H_\text{b}$, $\rho(0) = \ket{n_1, \dots, n_N}$ with $n_i$ if $d_i>0$ and zero otherwise. The state is fully characterized by the covariance matrix $\Gamma = \diag(1-\mathcal{N}, \mathcal{N})$ with $\mathcal{N}=\diag(n_1,\dots,n_N)$. Let $\mathcal{G} = VWV^\dagger$ be the diagonalization of $\mathcal{G}$ with $W=\diag(\omega_1,\dots,\omega_{2N})$ and $V^\dagger V=1$. Define the quasiparticle operators $\gamma = V^\dagger\Psi$, which evolve as $\gamma_a(t) = e^{-i\omega_at}\gamma_a(0)$. Substituting $\Psi=V\gamma$ in~\eqlabel{eq:c}, we have that
\begin{equation}
    [H_\text{b}, H_\text{c}] = \frac{1}{2}\gamma^\dagger V^\dagger\mathcal{C}V\gamma = \frac{1}{2}\gamma^\dagger\tilde{\mathcal{C}}\gamma \implies e^{-iH_\text{c}t} [H_\text{b}, H_\text{c}] e^{iH_\text{c}t} = \frac{1}{2} \sum_{a,b=1}^{2N} \gamma^\dagger_a \tilde{\mathcal{C}}(t) \gamma_b \quad\text{with }\tilde{\mathcal{C}}(t)=\tilde{\mathcal{C}}e^{i(\omega_a-\omega_b)t}.
\end{equation}
Redefining the covariance matrix in the quasiparticle basis as $K=V^\dagger\Gamma V$, we finally obtain that $P(t) = -\frac{i}{2}\trace{K\tilde{\mathcal{C}}(t)}$.%

For completeness, now we compute the exact powers for a generic $n_L=2$ setup. We take as battery Hamiltonian $H_\text{b} = (\omega_0/2) \sum_{i=1}^N (1 + \sigma^z_i)$, its ground state as initial state $\rho(0) = \ket{\downarrow}\!\bra{\downarrow}$ and as charger Hamiltonian the anisotropic XY chain under periodic boundary conditions (PBC) with a transverse field as most general model, namely
\begin{equation}\label{eq:xy_chain}
    H_\text{c} = \sum_{\braket{ij}} \left( \frac{g+\Delta}{2}\sigma^x_i\sigma^x_j + \frac{g-\Delta}{2}\sigma^y_i\sigma^y_j \right) + b\sum_{i=1}^N \sigma^z_i.
\end{equation}
Here, $g$ is the hopping strength, $\Delta$ the anisotropy, $b$ the transverse field and $\braket{\cdot}$ denotes that enclosed qubit indices
are nearest neighbours with $\sigma^\alpha_{N+1} = \sigma^\alpha_1$. 
Using the Jordan-Wigner transformation [\eqlabel{eq:jwt}], it becomes
\begin{equation}\label{eq:xy_chain_jwt}
    H_\text{c} = \sum_{\braket{ij}} \left( gc^\dagger_i c_j - \Delta c_i c_j +\hc \right) + b\sum_{i=1}^N (1-2c^\dagger_i c_i).
\end{equation}
For the even (odd) parity sector, we use antiperiodic (periodic) boundary conditions $c_{N+1}=-c_1$ ($c_{N+1}=c_1$). Next, we apply the Fourier transform $c_l = (e^{-i\pi/4}/\sqrt{N}) \sum_{k\in K_\text{BZ}} e^{ik(la)} c_k$, with $a$ the lattice spacing and momenta $K_\text{BZ} = \{(2i-1)\pi/N \,|\,i=1,\dots,N\}$ ($K_\text{BZ} = \{2i\pi/N\,|\,i=0,\dots,N-1\}$) for even (odd) parity sectors, consistent with the boundary Jordan-Wigner boundary conditions. Doing so,~\eqlabel{eq:xy_chain_jwt} reads in momentum space as%\vspace{-0.5mm}
\begin{equation}
    H_\text{c} = \sum_{k\in K_\text{BZ}} \left\{ 2[g\cos(ka) - b]c^\dagger_k c_k - \Delta\sin(ka)( c^\dagger_k c^\dagger_{-k} + c_{-k} c_k ) \right\} + \text{const.} = \frac{1}{2}\sum_{k\in K_\text{BZ}} \Psi^\dagger_k H_\text{c}(k) \Psi_k + \text{const.},
\end{equation}
with $\Psi_k = [c_k, c^\dagger_{-k}]^\text{T}$ and $H_\text{c}(k) = 2[g\cos(ka) - b]\sigma^z - 2\Delta\sin(ka)\sigma^x$ with eigenvalues $\pm 2E_k$, where $E_k^2 = [g\cos(ka) - b]^2 + \Delta^2\sin^2(ka)$, and $\tau^\alpha$ the Pauli matrices with $\alpha=x,y,z$. This system satisfies the Schrödinger-like equation \smash{$i\dot{U}_k(t) = H_\text{c}(k) U_k(t)$}, with $U_k(t)$ the time-evolution operator in momentum space. Its solution reads as
\begin{equation}
\begin{split}
    U_k(t) &= \mathfrak{T}\exp\left[ -i\int_0^t \text{d}t' H_\text{c}(k) \right] = \exp\left[ -2it\left( [g\cos(ka) - b]\tau^z - \Delta\sin(ka)\tau^x \right) \right] \\
    &= \cos(2E_kt) - i\frac{\sin(2E_kt)}{E_k}\left( [g\cos(ka) - b]\tau^z - \Delta\sin(ka)\tau^x \right) = \begin{bmatrix} \mathfrak{a}_k(t) & -\mathfrak{b}^*_k(t) \\ \mathfrak{b}_k(t) & \mathfrak{a}^*_k(t) \end{bmatrix},
\end{split} 
\end{equation}
where $\mathfrak{T}\exp[\cdot]$ denotes the time-ordered exponential and we use the identity $\exp(-i\theta(\hat{n}\cdot\vec{\sigma})) = \cos\theta -i\sin\theta(\hat{n}\cdot\vec{\sigma})$.%

The state evolves as $\rho_k(t) = \ket{\psi_k(t)}\!\bra{\psi_k(t)} = U_k(t)\rho_k(0)U^\dagger_k(t)$, with $\rho_k(0) = \ket{\downarrow}\!\bra{\downarrow}$ and $\ket{\psi_k(t)} = U_k(t)\ket{\downarrow} = [u_k(t),v_k(t)]^\text{T}$ with $|u_k(t)|^2 + |v_k(t)|^2=1$ and
\begin{align}
    u_k(t) &= -\mathfrak{b}^*_k(t) = \frac{i\Delta\sin(ka)\sin(2E_kt)}{E_k}, \qquad
    v_k(t) = \mathfrak{a}^*_k(t) = \cos(2E_kt) +i\frac{[g\cos(ka) - b]\sin(2E_kt)}{E_k}.
\end{align}
Therefore, the energy injected is given by
\begin{equation}
    E(t) = \frac{\omega_0}{2}\sum_{k\in K_\text{BZ}} \braket{\psi_k(t) | 1+\sigma^z | \psi_k(t) } = \frac{\omega_0}{2}\left[ N + \sum_{k\in K_\text{BZ}} (2|u_k(t)|^2 - 1) \right] = \omega_0 \Delta^2\sum_{k\in K_\text{BZ}} \frac{\sin^2 (ka)}{E_k^2}\sin^2 (2E_kt),
\end{equation}
returning as average and instantaneous charging powers $P_\text{avg}(t) = E(t)/t$ and
\begin{equation}
    P(t) = \trace{\dot{\rho}(t)H_\text{b}} = 2\omega_0 \Delta^2\sum_{k\in K_\text{BZ}} \frac{\sin^2 (ka)}{E_k}\sin(4E_kt),
\end{equation}
respectively, which clearly scale extensively. Thermodynamic limit can be taken by replacing $\sum_{k\in K_\text{BZ}} \mapsto (N/2\pi)\int_0^{2\pi} \text{d}k$. 

As an example, for the particular case of $g=b=0$, the energy injected simplifies to
\begin{equation}\label{eq:energy_xy}
    E(t) = \omega_0 \sum_{k\in K_\text{BZ}} \sin^2(2t|\Delta\sin(ka)|) \implies \lim_{N\to\infty} \frac{E(t)}{N} = \frac{\omega_0}{2}\left[ 1 - J_0(4|\Delta|t) \right],
\end{equation}
with $J_\alpha(x)$ the Bessel function of the first kind of order $\alpha$, and the instantaneous charging power reads as
\begin{equation}\label{eq:power}
    P(t) = 2\omega_0 |\Delta| \sum_{k\in K_\text{BZ}} |\sin(ka)|\sin(4t|\Delta\sin(ka)|) \implies \lim_{N\to\infty} \frac{P(t)}{N} = 2\omega_0 |\Delta| J_1(4|\Delta|t).
\end{equation}
\begin{figure}[!tb]
    \centering
    \includegraphics[width=.97\linewidth]{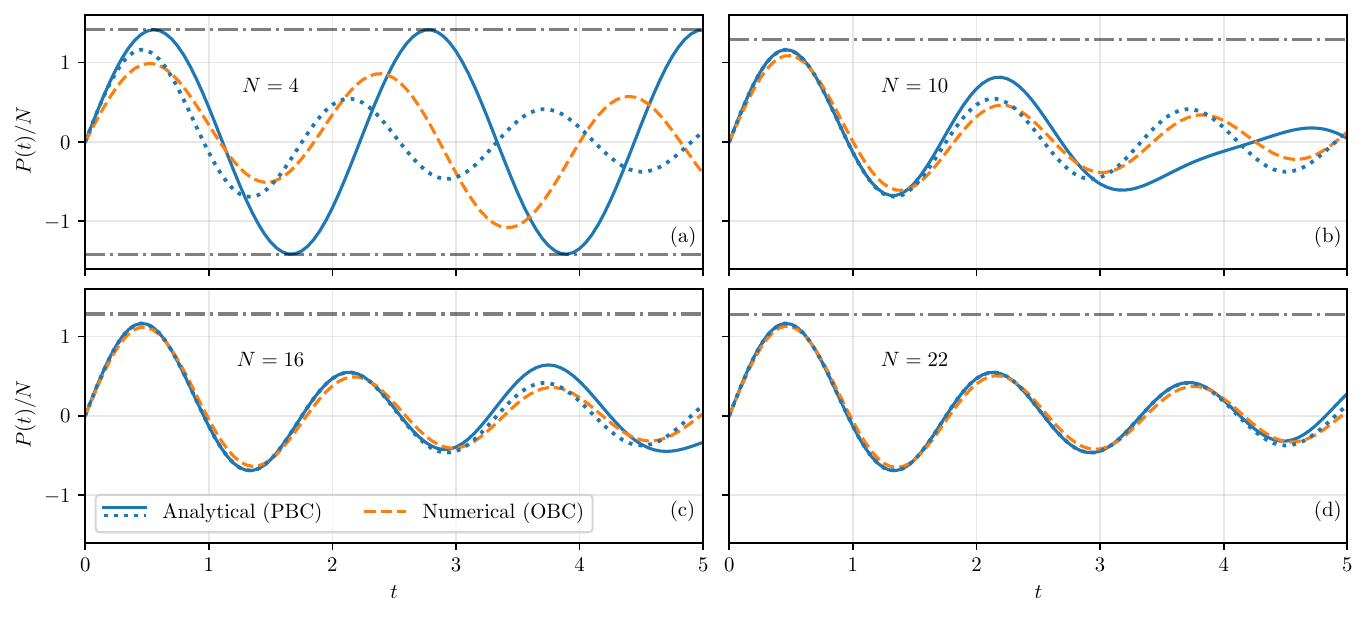}\vspace{-2mm}%
    \caption{\textbf{Comparison of instantaneous charging powers.} Using~\eqlabel{eq:xy_chain} as charger, the normalized charging powers are computed for $N=4,10,16,22$ (\textbf{a}, \textbf{b}, \textbf{c}, and \textbf{d}, respectively) using the analytical finite-$N$ and thermodynamic-limit results of~\eqlabel{eq:power} and the exact numerical computation (solid, dotted and dashed lines, respectively). The exact operator norm of $[H_\text{b}, H_\text{c}]$ is included as a reference (dash-dotted, \eqlabel{eq:exact_commutator_norm}). The parameters used are $g=b=0$ and $\Delta=\omega_0=1$.%
    }\label{fig:quadratic}\vspace{-2mm}
\end{figure}%

For completeness, we can also compute analytically the largest singular value of $[H_\text{b},H_\text{c}]$ to evaluate the tightness of the $|P(t)|\le \norm{[H_\text{b},H_\text{c}]}$ bound for integrable cases. Following the same procedure as before, we obtain that
\begin{equation}
    [H_\text{b}, H_\text{c}] = i\omega_0\Delta\sum_{\braket{ij}} (\sigma^y_i\sigma^x_j + \sigma^x_i\sigma^y_j) = -2\omega_0\Delta\sum_{\braket{ij}}(c^\dagger_ic^\dagger_j - \hc) = 2\omega_0\Delta \sum_{k\in K_\text{BZ}} \sin(ka)( c^\dagger_kc^\dagger_{-k} - c_{-k}c_k ).
\end{equation}
Let $M_k= c^\dagger_kc^\dagger_{-k} - c_{-k}c_k$. Since $M^2_k=-1$ in the even parity sector $(k,-k)$ of the full four-dimensional Fock space, their eigenvalues are $\pm i$. In the odd sector the two corresponding eigenvalues are zero, thus we finally obtain that
\begin{equation}\label{eq:exact_commutator_norm}
    \norm{[H_\text{b}, H_\text{c}]} = 2|\omega_0\Delta|\sum_{k\in K_\text{BZ}} |\sin(ka)| = 2|\omega_0\Delta| \cdot \begin{cases}
        \cot \frac{\pi}{2N} & \text{if }N\text{ odd} \\
        2\csc\frac{\pi}{N} & \text{otherwise}
    \end{cases} \implies \lim_{N\to\infty} \frac{\norm{[H_\text{b}, H_\text{c}]}}{N} = \frac{4}{\pi}|\omega_0\Delta|,
\end{equation}
where we use the finite arithmetic progression identity $\sum_{m=0}^M \sin(a+mx) = \sin(a+Mx/2)\sin[(M+1)x/2]/\sin(x/2)$.%

Despite PBC are considered throughout the derivation, PBC results can also be leveraged for open boundary conditions (OBC) up to an error of order $\mathcal{O}[1/N]$. Setting as lattice spacing $a=1$, in~\figlabel{fig:quadratic} we compare for $N= 4, 10, 16, 22$, the normalized powers $P(t)/N$ for an open chain described by~\eqlabel{eq:xy_chain} with $(g,\Delta,b)=(0,1,0)$ and $\omega_0=1$ for $H_\text{b}$, which are computed numerically, with respect to the analytical results (finite-$N$ and thermodynamic limit) but considering PBC. The $\norm{[H_\text{b}, H_\text{c}]}$ norm is plotted as a reference [\eqlabel{eq:exact_commutator_norm}]. As expected: i) charging power scales extensively, and ii) both of the analytical results converge to the numerical one as $N$ increases.

\section{Expectation value of participation ratios and large-\texorpdfstring{$N$}{N} asymptotics}

We propose $N_q=(\sum_{k=1}^N p^q_k)^{1/(1-q)}$ as participation number, with $p_k= \sum_{s\in S^{(k)}} |c^{(k)}_s| / \sum_{k=1}^N \sum_{s\in S^{(k)}} |c^{(k)}_s|$ the relative contribution of the $k$th cell of $H_\text{b}$, $[h_k,H_\text{c}(t)] = \sum_{s\in S^{(k)}} c^{(k)}_s P_s$ the Pauli decomposition and $S^{(k)}$ the set of Pauli strings present. The weights $p_k$ can be computed analytically by accounting for the diagonal contribution
\begin{equation}\label{eq:pk_diag}
\begin{split}
    &[\delta_k\sigma^z_k, H_\text{c}] = i\delta_k \sum^{k-1}_{l=1} \Big[ \I{A_{lk} - B_{lk}} \sigma^x_l \St{l}{k} \sigma^x_k + \R{A_{lk} + B_{lk}} \sigma^x_l \St{l}{k} \sigma^y_k - \R{A_{lk} - B_{lk}} \sigma^y_l \St{l}{k} \sigma^x_k + \I{A_{lk} + B_{lk}} \sigma^y_l \St{l}{k} \sigma^y_k \Big] \\
    &\qquad-i\delta_k \sum^N_{l=k+1} \!\!\Big[ \I{A_{kl} + B_{kl}} \sigma^x_k \St{k}{l} \sigma^x_l + \R{A_{kl} - B_{kl}} \sigma^x_k \St{k}{l} \sigma^y_l - \R{A_{kl} + B_{kl}} \sigma^y_k \St{k}{l} \sigma^x_l + \I{A_{kl} - B_{kl}} \sigma^y_k \St{k}{l} \sigma^y_l  \Big],
\end{split}
\end{equation} 
as well as the off-diagonal contributions, which read as
\begin{equation}\label{eq:pk_offdiag}
\begin{split}
    &[\beta_k\sigma^x_k + \gamma_k\sigma^y_k, H_\text{c}] = -iA_{kk}(\gamma_k\sigma^x_k - \beta_k\sigma^y_k) \\
    &\qquad-i\sum^{k-1}_{l=1}\Big[ (\beta_k \I{A_{lk} - B_{lk}} - \gamma_k \R{A_{lk} + B_{lk}} ) \sigma^x_l\St{l,}{k+1} - (\beta_k \R{A_{lk} - B_{lk}} - \gamma_k \I{A_{lk} + B_{lk}} ) \sigma^y_l\St{l,}{k+1} \Big] \\
    &\qquad+i\sum^N_{l=k+1}\Big[ (\beta_k\I{A_{kl} + B_{kl}} - \gamma_k\R{A_{kl} + B_{kl}} ) \St{k-1,}{l}\sigma^x_l + (\beta_k\R{A_{kl} - B_{kl}} + \gamma_k \I{A_{kl} - B_{kl}} ) \St{k-1,}{l}\sigma^y_l \Big] \\
    &\qquad-i\sum^{k-1}_{l=1}\sum^N_{m=k+1} \R{A_{lm} + B_{lm}} \left( \beta_k \sigma^x_l \St{l}{k} \sigma^y_k \St{k}{m} \sigma^x_m - \gamma_k \sigma^x_l \St{l}{k} \sigma^x_k \St{k}{m} \sigma^x_m \right) \\
    &\qquad-i\sum^{k-1}_{l=1}\sum^N_{m=k+1} \R{A_{lm} - B_{lm}} \left( \beta_k \sigma^y_l \St{l}{k} \sigma^y_k \St{k}{m} \sigma^y_m - \gamma_k \sigma^y_l \St{l}{k} \sigma^x_k \St{k}{m} \sigma^y_m \right) \\
    &\qquad+i\sum^{k-1}_{l=1}\sum^N_{m=k+1} \I{A_{lm} - B_{lm}} \left( \beta_k \sigma^x_l \St{l}{k} \sigma^y_k \St{k}{m} \sigma^y_m - \gamma_k \sigma^y_l \St{l}{k} \sigma^x_k \St{k}{m} \sigma^y_m \right) \\
    &\qquad-i\sum^{k-1}_{l=1}\sum^N_{m=k+1} \I{A_{lm} + B_{lm}} \left( \beta_k \sigma^y_l \St{l}{k} \sigma^y_k \St{k}{m} \sigma^x_m - \gamma_k \sigma^y_l \St{l}{k} \sigma^x_k \St{k}{m} \sigma^x_m \right).
\end{split}
\end{equation}
with equations~\eqref{eq:pk_diag} and~\eqref{eq:pk_offdiag} returning $4(N-1)$ and $2[N+4(k-1)(N-k)]$ different Pauli strings, respectively. 
To finally compute the relative weights $p_k$, one has to sum over the absolute values of each of the Pauli coefficients, where the total number scales in general as $N_{\max} = \sum^N_{k=1} N^{(k)}_{\max} = \sum^N_{k=1} \big[ 2+6(N-1) + 8(k-1)(N-k) \big] = 2N(N+2)(2N-1)/3\sim\mathcal{O}[N^3]$, value that is used to normalize $N_q$ and directly correlate it, in the main text, with the cumulative nonlocality number $n_L$ and power scaling exponents associated.

\emph{Exact ensemble average for finite system sizes and arbitrary locality cutoff.}---At finite-$N$, the ensemble average of the participation ratio can be computed as
\begin{equation}\label{eq:participation_number}
    \mathbb{E}[\eta_N(n_L)] = \frac{1}{N_{\max}} \sum_{k=1}^N \mathbb{E}[N_1^{(k)}] = \frac{1}{N_{\max}} \sum_{k=1}^N \mathbb{E}[\exp(H_k)],
\end{equation}
where $H_k$ is the Shannon entropy of the $p_s^{(k)}$ normalized weights for the $k$th commutator, which reads as 
\begin{equation}
    H_k = \log S_k -\frac{R_k}{S_k} \implies \exp(H_k) = S_k\exp\left( -\frac{R_k}{S_k} \right), \qquad\text{with }\begin{cases}
    S_k = \sum_{s\in S^{(k)}} |c_s^{(k)}| \\ R_k = \sum_{s\in S^{(k)}} |c_s^{(k)}| \log |c_s^{(k)}|
    \end{cases}.
\end{equation}
Its expectation value is computed over random matrices with independent real and imaginary Gaussian variables. Since their normalization cancels in $H_k$, we can treat all variables as independent normal.%

For fixed $k$, only matrix matrix elements within a distance $n_L-1$ of $k$ participate. Let $a_k=\min(k-1, n_L-1)$ and $b_k=\min(N-k,n_L-1)$ count the allowed sites to the left and right of $k$, respectively. Furthermore, a pair $(l,m)$ with $l<k<m$ is allowed if $m-l\le n_L-1$, with $T_k=\sum_{l=\max(1, k-n_L+2)}^{k-1}\sum_{m=k+1}^{\min(N, l+n_L-1)} 1$ the number of such pairs. Then, the $k$th commutator has $d_k=4(1+a_k+b_k+T_k)$ Gaussian variables, decomposed as%
\begin{itemize}
    \item The first four variables come from the battery Hamiltonian coefficients and diagonal of $A$: $\beta_k$, $\gamma_k$, $\delta_k$ and $A_{kk}$.
    \item Every matrix entry contribute with four independent random Gaussian variables: $\R{A\pm B}$ and $\I{A\pm B}$.
\end{itemize}%

The goal is to compute, for each commutator $k$, the expectation value
\begin{equation}\label{eq:gaussian_exp}
    I_k = \mathbb{E}[\exp(H_k)] = \int_{\mathbb{R}^{d_k}} \text{d}z \exp[H_k(z)]\Phi_{d_k}(z),
\end{equation}
with $\Phi_{d_k}(z)$ the standard $d_k$-dimensional Gaussian density function and $z=[z_1,\dots,z_{d_k}]$ the reduced Gaussian variables. This integral can be evaluated by quasi-Monte Carlo methods~\cite{lemieux2009quasi-montecarlo}. We transform~\eqlabel{eq:gaussian_exp} into an integral over the unit hypercube $[0,1]^{d_k}$. Let $\Phi(t)=(1/\sqrt{2\pi})\int_{-\infty}^{t} \text{d}x e^{-x^2/2}$ be the Gaussian cumulative density and set $u_j = \Phi(z_j)$, then $u_j\in[0,1]$ with Jacobian $\text{d}z_j/\text{d}u_j = 1/\Phi(z_j)$. Substituting into~\eqlabel{eq:gaussian_exp}, we obtain%
\begin{equation}\label{eq:sobol_exp}
    I_k = \int_{[0,1]^{d_k}} \text{d}u_1 \dots \text{d}u_{d_k} \exp[H_k(\Phi^{-1}(z_1), \dots, \Phi^{-1}(z_{d_k}))].
\end{equation}
Ordinary Monte Carlo methods sample $u_i$ independently and uniformly from $[0,1]^{d_k}$, with convergence rate $\mathcal{O}[M^{-1/2}]$, where $M$ is the number of samples. Quasi-Monte Carlo methods instead uses a deterministic low-discrepancy sequence that fills the unit hypercube more evenly, with convergence rate $\mathcal{O}[(\log M)^d/M]$ for sufficiently smooth integrands. This is advantageous for moderate dimensions $d$. Here, we choose Owen-scrambled Sobol' sequences~\cite{lemieux2009quasi-montecarlo} to obtain a conventional error estimate, since pure Sobol' sequences are deterministic. 
Then, the quasi-Monte Carlo estimate is obtained from $S$ independent Owen-scrambled Sobol' sequences, each containing $M=2^m$ points. Let $z^{(i,s)}$ denote the $i$th Gaussian sample for the $s$th scramble. The mean and variance of the mean estimates are given by%
\begin{equation}
    \bar{I}_k = \frac{1}{S} \sum_{s=1}^{S} \frac{1}{M} \sum_{i=1}^M \exp[H_k(z^{(i,s)})], \qquad s^2(\bar{I}_k) = \frac{1}{S(S-1)} \sum_{s=1}^S \bigg( \frac{1}{M} \sum_{i=1}^M \exp[H_k(z^{(i,s)})] - \bar{I}_k \bigg)^2.
\end{equation}

\emph{Exact $n_L=1$ participation ratio.}---A particular case arises when the locality cutoff $n_L$ is set to one (local charger limit), where only two Pauli strings are present in the $k$th commutator, with corresponding coefficients given by $\kappa_{1,1}^{(k)} = -A_{kk}\gamma_k$ and $\kappa_{1,2}^{(k)} = A_{kk}\beta_k$ [\eqlabel{eq:c1}]. Its relevant random variables involved are $A_{kk}$, $\beta_k$ and $\gamma_k$, which are independent real Gaussian variables. Therefore, using $r=|\beta_k|/|\gamma_k|$, the normalized weights are given by
\begin{equation}
    p_1^{(k)} = \frac{|A_{kk}||\gamma_k|}{|A_{kk}||\gamma_k| + |A_{kk}||\beta_k|} = \frac{1}{1+r}, \qquad p_2^{(k)} = \frac{|A_{kk}||\beta_k|}{|A_{kk}||\gamma_k| + |A_{kk}||\beta_k|} = \frac{r}{1+r}.
\end{equation}
Consequently,
\begin{equation}
    \sum_{s\in S^{(k)}} p_s^{(k)} \log p_s^{(k)} = \frac{1}{1+r}\log \frac{1}{1+r} + \frac{r}{1+r}\log \frac{r}{1+r} = \frac{r}{1+r}\log r -\log(1+r) \implies \eta^{(k)} = \frac{1+r}{2}r^{-r/(1+r)}.
\end{equation}

Since the absolute ratio of two normal variables is half-Cauchy~\cite{forbes2011normal,gelman2006prior}, with probability density function $f(r) = 2/[\pi(1+r^2)]$ for $r\ge 0$, the expectation value of $\eta^{(k)}$ for $n_L=1$ can be obtained as
\begin{equation}
    \eta_{n_L=1} = \mathbb{E}[\eta^{(k)}] = \int_0^\infty \text{d}r \eta^{(k)}(r)f(r) = \int_0^\infty \text{d}r \frac{1+r}{2}r^{-r/(1+r)} \frac{2}{\pi(1+r^2)} \approx 0.8671.
    % https://www.wolframalpha.com/input?i2d=true&i=Divide%5B1%2C%CF%80%5DIntegrate%5BDivide%5B1%2Br%2C1%2BPower%5Br%2C2%5D%5DPower%5Br%2C-Divide%5Br%2C1%2Br%5D%5D%2C%7Br%2C0%2C%E2%88%9E%7D%5D
\end{equation}
Summing over all commutators, for $n_L=1$ we have that $\sum_{k=1}^N N^{(k)}_{\max} = 2N$, returning that
\begin{equation}\label{eq:asymptotic_nL1}
    \mathbb{E}[\eta_N(n_L=1)] = \frac{2N}{N_{\max}} \eta_{n_L=1} = \frac{3\eta_{n_L=1}}{(N+2)(2N-1)} = \frac{3\eta_{n_L=1}}{2N^2} - \frac{9\eta_{n_L=1}}{4N^3} + \mathcal{O}[N^{-4}],
\end{equation}
which for large $N$, it decays quadratically with increasing system size, as expected.

\emph{Exact participation ratio for $n_L=2$.}---For $n_L=2$, $C_2=0$. Therefore, so only local and two-body contributions from $C_1$, $C_3$ and $C_4$ families matter. We split the two-body terms into edge and bulk contributions depending on the quantum cell index, where: $k=1$ has only a right neighbour, $k=N$ has only a left neighbour and $1<k<N$ has two at both sides. Hence there are two edge commutators and $N-2$ bulk commutators. We define $D_{(\text{bulk}, \text{edge})} = \mathbb{E}[N_{(\text{bulk}, \text{edge})}^{(k)}]$. The distributions are independent of the absolute position of $k$, so
\begin{equation}\label{eq:exact_popratio_nL2}
    \sum_{k=1}^N \mathbb{E}[N_1^{(k)}] = (N-2)D_\text{bulk} + 2D_\text{edge} \implies \mathbb{E}[\eta_N(n_L=2)] = \frac{3[(N-2)D_\text{bulk} + 2D_\text{edge}]}{2N(N+2)(2N-1)},
\end{equation} 
which is an exact finite-$N$ expression. The next step is to determine the $D_{(\text{bulk}, \text{edge})}$ expressions analytically.%

Bulk commutators have exactly $2+12(n_L-1)=14$ coefficients. Let $\beta=\beta_k$, $\gamma=\gamma_k$, $\bm{\delta} = (\delta_-, \delta_0, \delta_+) = (\delta_{k-1}, \delta_k, \delta_{k+1})$ and $A=A_{kk}$. In polar coordinates $\beta=r\cos\theta$ and $\gamma=r\sin\theta$, whose joint distribution becomes $p(\beta,\gamma) = (1/2\pi)e^{-(\beta^2+\gamma^2)/2} = (r/2\pi)e^{-r^2/2}$ with $r^2=\beta^2+\gamma^2$ and Jacobian of the transformation $\text{d}\beta\text{d}\gamma = r\text{d}r\text{d}\theta$ and $(r,\theta)\in [0,\infty)\times[0,2\pi)$. Dividing the $14$ coefficients by $r$ removes the $r$ dependency from the Shannon entropy.%

Now, let $\bm{q} = (q_-, q_0, q_+) = \bm{\delta}/r$, with Jacobian of the transformation $\text{d}^3\delta = r^3\text{d}^3q$ and joint density function 
\begin{equation}
    p(r, \bm{q}) = \frac{r^4}{(2\pi)^{3/2}}\exp\left[ -\frac{r^2}{2} \big( 1 + \norm{\bm{q}}^2 \big) \right] \qquad\text{with }\norm{\bm{q}}^2 = q_-^2 + q_0^2 + q_+^2.
\end{equation}
Integrating over $r$ and using that $\int_0^\infty \text{d}r r^4e^{-ar^2/2} = 3\sqrt{2\pi}/2a^{5/2}$, we finally obtain that
\begin{equation}
    p_3(\bm{q}) = \frac{3}{4\pi}\frac{1}{\big( 1+\norm{\bm{q}}^2 \big)^{5/2}}.
\end{equation}
Therefore, the variable $r$ has been analytically integrated away, thus the final integral has $13$ instead of $14$ variables. Since $A=|A_{kk}|$ is half-normal, with probability density function $\varphi(x) = \sqrt{2/\pi}e^{-x^2/2}$ ($x\ge 0$), we obtain
\begin{equation}
\begin{split}
    D_\text{bulk} &= \frac{1}{2\pi} \int_0^{2\pi} \text{d}\theta \sqrt{\frac{2}{\pi}} \int_0^\infty \text{d}A e^{-A^2/2} \frac{3}{4\pi} \int_{\mathbb{R}^3} \frac{\text{d}^3q}{\big( 1 + \norm{\bm{q}}^2 \big)^{5/2}} \int_{\mathbb{R}^8} \frac{\text{d}^8z}{(2\pi)^4} \exp[H_k(\theta,A,\bm{q},\bm{z})]. 
\end{split}
\end{equation}

Same logic applies for edge commutators. Let $\bm{\delta} = (\delta_\text{l},\delta_\text{r}) = (\delta_1, \delta_N)$ and $\bm{q} = (q_\text{l}, q_\text{r}) = \bm{\delta}/r$ with joint density 
\begin{equation}
    p_2(\bm{q}) = \frac{1}{\pi}\frac{1}{\big( 1 + \norm{\bm{q}}^2 \big)^2} \qquad\text{with }\norm{\bm{q}}^2 = q_\text{l}^2 + q_\text{r}^2.
\end{equation}
Therefore, the integral form of $D_\text{edge}$ becomes an eight-dimensional integral of the form
\begin{equation}
    D_\text{edge} = \frac{1}{2\pi} \int_0^{2\pi} \text{d}\theta \sqrt{\frac{2}{\pi}} \int_0^\infty \text{d}A e^{-A^2/2} \frac{1}{\pi} \int_{\mathbb{R}^2} \frac{\text{d}^2q}{\big( 1 + \norm{\bm{q}}^2 \big)^{2}} \int_{\mathbb{R}^4} \frac{\text{d}^4z}{(2\pi)^2} \exp[H_k(\theta,A,\bm{q},\bm{z})].
\end{equation}

We can expand in powers of $N$ the exact finite-$N$, $n_L=2$ result obtained in~\eqlabel{eq:exact_popratio_nL2}, where
\begin{equation}\label{eq:asymptotic_nL2}
    \mathbb{E}[\eta_N(n_L=2)] = \frac{3[(N-2)D_\text{bulk} + 2D_\text{edge}]}{2N(N+2)(2N-1)} = \frac{3D_\text{bulk}}{4N^2} + \frac{3}{4N^3} \left( 2D_\text{edge} - \frac{7}{2}D_\text{bulk} \right) + \mathcal{O}[N^{-4}],
\end{equation}
which confirms the leading $N^{-2}$-asymptotics with $D_\text{bulk}\approx 9.9144$ and $D_\text{edge}\approx 5.8177$ after numerical evaluation.

\emph{Asymptotic behaviour for $n_L\ge3$ and large $N$.}---As observed in Fig. 3a in the main text, the participation ratio for the dense case ($n_L=N$) presents a plateau around $0.6-0.7$, consequence of the normal distributions used to define $H_\text{b}$ and $H_\text{c}$ Hamiltonians. We can compute this quantity as%
\begin{equation}
    \eta_\text{crit} = \lim_{N\to\infty} \frac{\sum_{k=1}^N N^{(k)}_1}{N_{\max}} = \lim_{N\to\infty} \frac{\sum_{k=1}^N N^{(k)}_{\max} \eta^{(k)}}{\sum_{k=1}^N N^{(k)}_{\max}} = \mathbb{E}[\eta]=\text{const.}, \qquad \text{with }\eta^{(k)} = \frac{N^{(k)}_1}{N^{(k)}_{\max}},
\end{equation}
constituting the large-$N$ limit of a weighted average of $\eta^{(k)}$, which reduces to computing the average value $\mathbb{E}[\eta]$~\cite{jamison1965convergence}.%

In the large-$N$ limit, equations~\eqref{eq:pk_diag}-\eqref{eq:pk_offdiag} show that the $C_2$-family contributions dominate, returning $8(i-1)(N-i)$ Pauli strings, while all other contributions are either constant or linear in $N$. Their coefficients are products of $\beta_i$ and $\gamma_i$ values times the sum or difference of the real and imaginary parts of $A_{kl}$ and $B_{kl}$, which are normally distributed. Hence these sums or differences are Gaussian with variance two. Since taking their absolute values yields half-normal variables~\cite{forbes2011normal}, so conditional on the $|\beta_i|$ and $|\gamma_i|$ magnitudes, the relevant scales are $\sigma_1=\sqrt{2}|\beta_i|$ and $\sigma_2=\sqrt{2}|\gamma_i|$.%

To compute the moments $\mu_X = \mathbb{E}[X]$ and $\mu_{X\log X} = \mathbb{E}[X\log X]$, with $X$ the mixture of two half-normal distributions $X_1$ and $X_2$ with scales $\sigma_1$ and $\sigma_2$, respectively, note that for a half-normal distribution $Y$ with scale $\sigma$ one obtains%
\begin{align} \label{eq:half_normal_x}
    \mu_Y &= \mathbb{E}[Y] = \frac{1}{\sigma}\sqrt{\frac{2}{\pi}} \int^\infty_0 ye^{-y^2/2\sigma^2} \text{d}y = \sigma\sqrt{\frac{2}{\pi}}, \\ \label{eq:half_normal_xlogx}
    \mu_{Y\log Y} &= \mathbb{E}[Y\log Y] = \frac{1}{\sigma}\sqrt{\frac{2}{\pi}} \int^\infty_0 y\log y e^{-y^2/2\sigma^2} \text{d}y = \sigma\sqrt{\frac{2}{\pi}} \left( \log\sigma + \frac{\log 2 - \gamma}{2} \right),
\end{align}
with $\gamma \approx 0.5772$ 
the Euler-Mascheroni constant. Since $X_1$ and $X_2$ contribute equally to $X$,  
$\mu_X$ and $\mu_{X\log X}$ read as
\begin{align}
    \mu_X = \frac{\mu_{X_1} + \mu_{X_2}}{2} = \frac{\sigma_1 + \sigma_2}{\sqrt{2\pi}}, \quad 
    \mu_{X\log X} = \frac{\mu_{X_1\log X_1} + \mu_{X_2\log X_2}}{2} = \frac{\sigma_1\log\sigma_1 + \sigma_2\log\sigma_2 + \frac{\log 2 - \gamma}{2}(\sigma_1 + \sigma_2)}{\sqrt{2\pi}}.
\end{align}

Therefore, introducing the ratio $r= \sigma_1/\sigma_2$, we obtain that
\begin{equation}\label{eq:eta_r}
    \eta = \exp \left( \log \mu_X - \frac{\mu_{X\log X}}{\mu_X} \right) = \frac{e^{\gamma/2}}{2\sqrt{\pi}}(\sigma_1+\sigma_2)\exp\left( -\frac{\sigma_1\log\sigma_1 + \sigma_2\log\sigma_2}{\sigma_1 + \sigma_2} \right) = %\frac{e^{\gamma/2}}{\sqrt{\pi}} \frac{1+r}{2}\sigma_2 \exp\left( -\frac{r\log r}{1+r} -\log\sigma_2 \right) = 
    \frac{e^{\gamma/2}}{\sqrt{\pi}} \frac{1+r}{2}r^{-r/(1+r)},
\end{equation}
constituting a function that only depends on the ratio of scales $r$. The quotient of two variables following half-normal distributions follows a half-Cauchy distribution, as aforementioned. Relabelling $\eta(r)\equiv\eta$, we finally obtain%
\begin{equation}\label{eq:eta_crit}
    \eta_\text{crit} = \mathbb{E}[\eta] = \int^\infty_0 \text{d}r \eta(r)f(r) = \frac{e^{\gamma/2}}{\sqrt{\pi}} \int^\infty_0 \text{d}r \frac{1+r}{2}r^{-r/(1+r)} \frac{2}{\pi(1+r^2)} \approx 0.6529.
    % https://www.wolframalpha.com/input?i2d=true&i=Divide%5BPower%5Be%2CDivide%5B%CE%B3%2C2%5D%5D%2CSqrt%5B%CF%80%5D%5DIntegrate%5BDivide%5Bx%2B1%2C2%5DPower%5Bx%2C-Divide%5Bx%2Cx%2B1%5D%5DDivide%5B2%2C%CF%80%5C%2840%291%2BPower%5Bx%2C2%5D%5C%2841%29%5D%2C%7Bx%2C0%2C%E2%88%9E%7D%5D
\end{equation}

From this result, we can derive for large $N$ and $n_L\ge 3$ an approximate formula of the participation ratio including the locality cutoff dependence explicitly.  
For arbitrary $n_L\ge 3$, the number of $C_2$-terms is given by $T_N(n_L) = 8\sum_{d=2}^{n_L-1}(N-d)(d-1) = 4(n_L-1)(n_L-2)(3N-2n_L)/3$. Under the $C_2$-family dominance assumption, we have
\begin{equation}\label{eq:asymptotic_nL3_exact}
    \mathbb{E}[\eta_N(n_L)] = \frac{\sum_{k=1}^N \mathbb{E}[N_1^{(k)}]}{N_{\max}} \approx \eta_\text{crit} \frac{T_N(n_L)}{N_{\max}} = 2\eta_\text{crit}\frac{(n_L-1)(n_L-2)(3N-2n_L)}{N(N+2)(2N-1)},
\end{equation}
with $\lim_{N\to\infty} 2(N-1)(N-2)/[(N+2)(2N-1)] = 1$, as expected. Now, for fixed $n_L$ and large $N$, note that
\begin{equation}\label{eq:asymptotic_nL3}
    \frac{2(3N-2n_L)}{N(N+2)(2N-1)} = \frac{3}{N^2}\left[ 1 - \frac{3}{2N} - \frac{2n_L}{3N} + \mathcal{O}[N^{-2}] \right] \implies \mathbb{E}[\eta_N(n_L)] \approx \frac{3\eta_\text{crit}(n_L-1)(n_L-2)}{N^2},
\end{equation}
which captures the $N^2$ decay present when $n_L\ll N$, observed in Fig. 3a in the main text.

In~\figlabel{fig:participation_exact}, we compute the ensemble-averaged participation ratios for $N\in[2,1000]$ and different locality cutoffs $n_L\in[1,100]$ via scrambled quasi-Monte Carlo evaluation of~\eqlabel{eq:gaussian_exp}. As reported in the main text, a plateau near $\eta_\text{crit}$ appears when $n_L\ge N$, thus when cumulative locality is present. For $n_L<N$, the participation ratios decrease quadratically, as shown in equations~\eqref{eq:asymptotic_nL1},~\eqref{eq:asymptotic_nL2}, and~\eqref{eq:asymptotic_nL3}. For $n_L\ge 3$ we also plot the asymptotic behaviour of the participation ratios through~\eqlabel{eq:asymptotic_nL3_exact}. As $n_L$ increases, these asymptotics converge toward the quasi-Monte Carlo results, validating our assumption that the $C_2$-family contribution dominates in this regime.
\begin{figure}[!tb]
    \centering
    \includegraphics[width=0.55\linewidth]{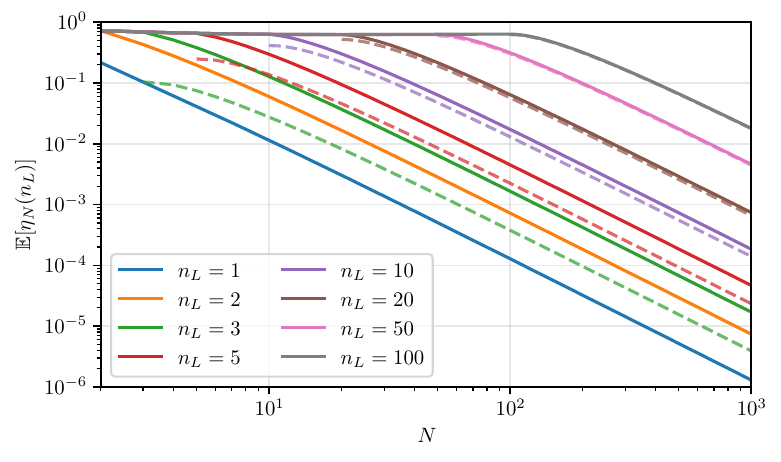}\vspace{-1mm}%
    \caption{\textbf{Expectation value of participation ratios}. Using scrambled quasi-Monte Carlo methods,~\eqlabel{eq:gaussian_exp} is evaluated for $N\in[2,1000]$ and $n_L\in[1, 100]$ with $2^{10}$ samples and eight Owen-scrambled Sobol' sequences~\cite{lemieux2009quasi-montecarlo} (solid lines). For $n_L\ge 3$ results, the asymptotic behaviour in~\eqlabel{eq:asymptotic_nL3_exact} is plotted as a reference (dashed). Error bars are not included as standard deviations are several orders of magnitude below the mean value for our configuration.
    }\label{fig:participation_exact}%
\end{figure}

\subsection{Power-law locality-decaying couplings}

Rather than considering constant entries for the $A$ and $B$ matrices in~\eqlabel{eq:quadratic}, we now take power-law decaying amplitudes in terms of their locality $L=l-k+1$, namely $\sigma_L = \sigma/L^p$, which describes the distance between the two fermion indices in $A_{kl}$ and $B_{kl}$ entries ($k\le l$). We then impose coupling strengths to decrease as $L^{-p}$, finding in $p=0$ the fully global $n_L=N$ case while in $p\to\infty$ we recover the $n_L=1$ local limit. Since only relative entry scales affect the expected participation ratios, renormalizing as in~\eqlabel{eq:renormalization} now yields%
\begin{equation}\label{eq:plaw_renormalization}
    \frac{1}{2^N}\braket{\trace{H_\text{c}^2(t)}} = \frac{\sigma^2}{2}\bigg( N + 2\sum_{L=2}^N \frac{N+1-L}{L^{2{p}}} \bigg) \implies \sigma^2 =  \frac{2N}{2(N+1)\zeta_N(2{p}) - 2\zeta_N(2{p}-1)-N}.
\end{equation}
Using~\eqlabel{eq:euler_maclaurin} in the large-$N$ limit, we get $2(1-p)(1-2p)N^{2p-1}$ if $0\le p<1/2$, $1/(\log N + \gamma - 3/2)$ if $p=1/2$, and $2/(2\zeta(2p)-1)$ if $p>1/2$. 
As expected, we obtain as limit cases $\sigma^2=2/N$ for $p=0$ and $\sigma^2=2$ for $p\to\infty$.

Note that the earlier $C_2$-dominance assumption holds at $p=0$ but not for general $p$. To see this, compare the $C_2$-per-cell contribution $W_k(p)$ with respected to the remaining families, $V_\text{local}$ and $V_k(p)$ for the $C_1$ and the combined $C_3$ and $C_4$ contributions, respectively. For $C_1$, $V_\text{local}=1$, while for the rest of families we have that
\begin{align}\label{eq:vkp}
    V_k(p) &= \sum_{l=1}^{k-1} \frac{1}{(k-l+1)^p} + \sum_{l=k+1}^N \frac{1}{(l-k+1)^p} = \sum_{L=2}^{a(k)+1} \frac{1}{L^p} + \sum_{L=2}^{b(k)+1} \frac{1}{L^p} = \zeta_{a(k)+1}(p) + \zeta_{b(k)+1}(p) - 2, \\ \label{eq:wkp}
    \begin{split}
        W_k(p) &= \sum_{L=3}^N \frac{1}{L^p} \sum_{l=\max(1,k-L+2)}^{\min(k-1, N-L+1)} 1 = \sum_{L=3}^N \frac{T_{k,L}}{L^p} = \sum_{L=3}^{a(k)+2} \frac{L-2}{L^p} + a(k)\sum_{L=a(k)+3}^{b(k)+2} \frac{1}{L^p} + \sum_{L=b(k)+3}^N \frac{N-L+1}{L^p} \\
    &= (N+1)\zeta_N(p) - [a(k)+2]\zeta_{a(k)+2}(p) - [b(k)+2]\zeta_{b(k)+2}(p) -\zeta_N(p-1) + \zeta_{a(k)+2}(p-1) + \zeta_{b(k)+2}(p-1) + 1.
    \end{split}
\end{align}
To obtain this, let $a(k)=\min(k-1,N-k)$, $b(k) = \max(k-1, N-k) = N-1-a(k)$, and $T_{k,L} = \min(k-1, N-L+1) - \max(1,k-L+2) + 1$ as the number of $(l,m)$ pairs with an interior $k$ index for a given $L$. It can be described as $T_{k,L} = L-2$ if $3\le L\le a(k)+2$, $a(k)$ if $a(k)+3 \le L\le b(k) + 2$ and $N-L+1$ otherwise. 

Following equations~\eqref{eq:pk_diag}-\eqref{eq:pk_offdiag}, we obtain as ratio $r_k(p)=8W_k(p) / ( 6V_k(p) + 2V_\text{local} ) = 4W_k(p) / ( 3V_k(p) + 1 )$. To compute its large-$N$ asymptotic behaviour, we introduce the continuum scaling $x=k/N$ ($0<x<1$), yielding the continuous interpolations $\mathcal{W}_N(x,p)=W_{k=xN}(p)$ and $\mathcal{V}_N(x,p) = V_{k=xN}(p)$. We also define $\alpha(x)=\min(x,1-x)$ and $\beta(x)=\max(x,1-x)=1-\alpha(x)$, which leads to $a(k)+2 = N\alpha(x)+c_1(x) = u(x)$ and $b(k)+2 = N\beta(x)+c_2(x) = v(x)$, with $(c_1(x), c_2(x)) = (1,2)$ if $x\le 1/2$ and $(2,1)$ otherwise. Substituting into equations~\eqref{eq:vkp}-\eqref{eq:wkp},
\begin{align}
    \mathcal{V}_N(x,p) &= H_{u(x)-1}(p) + H_{v(x)-1}(p) - 2, \\ \label{eq:wnxp}
    \mathcal{W}_N(x,p) &= (N+1)H_N(p) - u(x)H_{u(x)}(p) - v(x)H_{v(x)}(p) -H_N(p-1) + H_{u(x)}(p-1) + H_{v(x)}(p-1) + 1,
\end{align}
where $H_z(s) = \zeta(s) - \zeta(z+1,s)$ with $\zeta(z,s) = \sum_{m=0}^\infty (m+z)^{-s}$ the Hurwitz zeta function. Using~\eqlabel{eq:euler_maclaurin} in the large-$N$ limit, we can represent both expressions in terms of powers of $N$ as
\begin{align} \label{eq:wnxp_forms}
    \mathcal{W}_N(x,p) &= \begin{cases}
        N^{2-p}A_p(x) + N^{1-p}B_p(x) + D_p(x) + \mathcal{O}[N^{-p}] & \text{if }p\neq 1,2 \\
        NA_1(x) + B_1(x)\log N + D_1(x) + \mathcal{O}[N^{-1}] & \text{if }p=1 \\
        A_2(x)\log N + D_2(x) + \mathcal{O}[N^{-1}] & \text{if }p=2
    \end{cases}, \\
    \mathcal{V}_N(x,p) &= \begin{cases}
        N^{1-p}\tilde{A}_p(x) + \tilde{D}_p(x) + \mathcal{O}[N^{-p}] & \text{if }p\neq 1 \\
        \tilde{A}_1(x)\log N + \tilde{D}_1(x) + \mathcal{O}[N^{-1}] & \text{if }p=1
    \end{cases},
\end{align}
whose corresponding prefactors will be computed later. Taking their leading asymptotics per regime, we obtain that
\begin{equation}
    r_k(p) \mapsto \mathcal{r}_N(x,p) = \frac{ 4\mathcal{W}_N(x,p) }{ 3\mathcal{V}_N(x,p) + 1 } \sim \begin{cases}
        \mathcal{O}[N] & \text{if }p<1 \\
        \mathcal{O}[N/\log N] & \text{if }p=1 \\
        \mathcal{O}[N^{2-p}] & \text{if }1<p<2 \\
        \mathcal{O}[\log N] & \text{if }p=2 \\
        \mathcal{O}[1] & \text{if }p>2
    \end{cases}.
\end{equation}
Therefore, we expect $C_2$-family dominance only for $p\le 2$, and this dominance weakens as $p$ increases. Hence, large-$N$ analytic participation ratios are obtainable only for $p\le 2$ under this dominance. For $p>2$, all families contribute comparably, so correlations prevent the factorization leveraged when the $C_2$ family dominance is assumed and only finite-size calculations may be feasible. Nevertheless, we use the leading $C_2$-only behaviour to infer the asymptotic profile of the expected participation ratios including all $C_i$ contributions.

Assuming the $C_2$ family dominance, the $A_{lm}$ and $B_{lm}$ entries were previously sampled from a normal distribution and later rescaled globally according to~\eqlabel{eq:renormalization}. The power-law decay is now introduced by rescaling the entries as $a_{lm} = A_{lm}/L^p$ and $b_{lm} = B_{lm}/L^p$. Let $g_{1,2}^{(lm)} = \R{A_{lm}\pm B_{lm}}$ and $g_{3,4}^{(lm)} = \I{A_{lm} \pm B_{lm}}$. Therefore, for one pair $(l,m)$, the $k$th battery cell contributes to the $C_2$-family through the eight coefficients $|\beta_k||g_\alpha|/L^p$ and $|\gamma_k||g_\alpha|/L^p$ ($\alpha=1,2,3,4$). Then, the contribution of this pair to $S_k$ is $S_k^{(lm)} = (|\beta_k| + |\gamma_k|)L^{-p}\sum_{\alpha=1}^4 |g_\alpha|$. Therefore, conditional on the half-normal $|\beta_k|$ and $|\gamma_k|$, the expected value of $S_k^{lm}$ is given by
\begin{equation}
    \mathbb{E}[S_k^{(lm)} \,|\, |\beta_k|,|\gamma_k| ] = \frac{|\beta_k|+|\gamma_k|}{L^p} \sum_{\alpha=1}^4 \mathbb{E}[|g_\alpha|] = \frac{8( |\beta_k| + |\gamma_k| )}{\sqrt{\pi} L^p},
\end{equation}
where we use~\eqlabel{eq:half_normal_x}. Now, if we sum over the $(l,m)$ pairs, we have
\begin{equation}
\begin{split}
    \mathbb{E}[S_k \,|\, |\beta_k|,|\gamma_k|] &= \sum_{L=3}^N \sum_{l=\max(1,k-L+2)}^{\min(k-1, N-L+1)} \mathbb{E}[S_k^{(l,l+L-1)} \,|\, |\beta_k|,|\gamma_k|] = \frac{8(|\beta_k| + |\gamma_k|)}{\sqrt{\pi}} \sum_{L=3}^N \frac{T_{k,L}}{L^p} = \frac{8(|\beta_k| + |\gamma_k|)}{\sqrt{\pi}} W_k(p).
\end{split}
\end{equation}
Doing the same for $R_k$, we obtain
\begin{equation}
\begin{split}
    \mathbb{E}[R_k^{(lm)} \,|\, |\beta_k|,|\gamma_k| ] &= \frac{|\beta_k|}{L^p} \left[ \log\frac{|\beta_k|}{L^p} \sum_{\alpha=1}^4 \mathbb{E}[|g_\alpha \|] + \sum_{\alpha=1}^4 \mathbb{E}[|g_\alpha|\log|g_\alpha|] \right] + \frac{|\gamma_k|}{L^p} \left[ \log\frac{|\gamma_k|}{L^p} \sum_{\alpha=1}^4 \mathbb{E}[|g_\alpha \|] + \sum_{\alpha=1}^4 \mathbb{E}[|g_\alpha|\log|g_\alpha|] \right] \\
    &= \frac{8}{\sqrt{\pi}L^p} \left[ |\beta_k|\log|\beta_k| + |\gamma_k|\log|\gamma_k| + (|\beta_k| + |\gamma_k|)\left( \log 2 - \frac{\gamma}{2} -p\log L \right) \right].
\end{split}
\end{equation}
where we use~\eqlabel{eq:half_normal_xlogx}. Defining $Q_k(p) = \sum_{L=3}^N T_{k,L}\log L/L^p = -\partial_p W_k(p)$, summing over $(l,m)$ pairs brings
\begin{equation}
\begin{split}
     \mathbb{E}[R_k \,|\, |\beta_k|,|\gamma_k| ] &= \sum_{L=3}^N \sum_{l=\max(1,k-L+2)}^{\min(k-1, N-L+1)} \mathbb{E}[R_k^{(l,l+L-1)} \,|\, |\beta_k|,|\gamma_k|] \\
     &= \frac{8}{\sqrt{\pi}} \left\{ (|\beta_k|\log|\beta_k| + |\gamma_k|\log|\gamma_k|)W_k(p) + (|\beta_k|+|\gamma_k|)\left[ \left(\log 2 - \frac{\gamma}{2}\right) W_k(p) -pQ_k(p) \right] \right\} \\
     &= \mathbb{E}[S_k \,|\, |\beta_k|,|\gamma_k| ] \left( \frac{|\beta_k|\log|\beta_k| + |\gamma_k|\log|\gamma_k|}{|\beta_k| + |\gamma_k|} + \log2 -\frac{\gamma}{2} -p\frac{Q_k(p)}{W_k(p)}  \right).
\end{split}
\end{equation}
Introducing again the ratio $r=|\beta_k|/|\gamma_k|$ and using equations~\eqref{eq:participation_number} and~\eqref{eq:eta_r}, we obtain as expected participation number conditional on the half-Cauchy parameter $r$ that
\begin{equation}
    \mathbb{E}[\exp(H_k) \,|\, r] = \frac{4e^{\gamma/2}}{\sqrt{\pi}} (1+r)r^{-r/1+r} W_k(p) \exp\left[ p\frac{Q_k(p)}{W_k(p)} \right] = 8\eta(r) W_k(p) \exp\left[ p\frac{Q_k(p)}{W_k(p)} \right].
\end{equation}
Averaging over the half-Cauchy distribution as in~\eqlabel{eq:eta_crit}, we finally obtain as expected participation ratio 
\begin{equation}\label{eq:etanp}
    \mathbb{E}[\eta_N(p)] = \frac{1}{N_{\max}} \sum_{k=2}^{N-1} \mathbb{E}[\exp(H_k)] = \frac{8\eta_\text{crit}}{N_{\max}} \sum_{k=2}^{N-1} W_k(p) \exp\left[ p\frac{Q_k(p)}{W_k(p)} \right].
\end{equation}

Now we turn the sum over $k$ into an integral using that at the thermodynamic limit $\sum_{k=2}^{N-1}\mapsto (N-2)\int_0^1\text{d}x$. Taking $\mathcal{Q}_N(x,p) = -\partial_p \mathcal{W}_N(x,p)$, \eqlabel{eq:etanp} reduces into
\begin{equation}
    \mathbb{E}[\eta_N(p)] = \frac{8\eta_\text{crit}(N-2)}{N_{\max}} \int_0^1 \text{d}x \mathcal{W}_N(x,p) \exp\left[ p\frac{\mathcal{Q}_N(x,p)}{\mathcal{W}_N(x,p)} \right].
\end{equation}

Using the large-$N$ expansions of~\eqlabel{eq:wnxp}, we obtain for $p\neq 1,2$ as~\eqlabel{eq:wnxp_forms} prefactors that
\begin{align} \label{eq:apx}
    A_p(x) &= \frac{1 - \alpha^{2-p}(x) - \beta^{2-p}(x)}{1-p} - \frac{1 - \alpha^{2-p}(x) - \beta^{2-p}(x)}{2-p} = \frac{1 - x^{2-p} - (1-x)^{2-p}}{(1-p)(2-p)}, \\ \label{eq:bpx}
    B_p(x) &= \frac{1-c_1(x)\alpha^{1-p}(x)-c_2(x)\beta^{1-p}(x)}{1-p} = \frac{1-x^{1-p}-2(1-x)^{1-p}}{1-p}, \\ \label{eq:dpx}
    D_p(x) &= (1-c_1(x)-c_2(x))\zeta(p) - (1-1-1)\zeta(p-1) + 1 = 1 + \zeta(p-1) - 2\zeta(p).
\end{align}

Moving to the $p=1$ case, note that equations~\eqref{eq:apx}-\eqref{eq:bpx} presents a pole. 
Using that $z=Nc+d$ and large-$N$,
\begin{equation}\label{eq:zeta1}
    \zeta_{Nc+d}(1) = \log(Nc+d) +\gamma +\frac{1}{2(Nc+d)} + \mathcal{O}[N^{-2}] = \log N + \log c + \gamma + \left( d + \frac{1}{2} \right)\frac{1}{Nc} + \mathcal{O}[N^{-2}].
\end{equation}
we obtain as prefactors for $p=1$ in~\eqlabel{eq:wnxp_forms} that
\begin{align}\label{eq:a1x}
    A_1(x) &= (1-\alpha(x)-\beta(x))\gamma -\alpha(x)\log\alpha(x) -\beta(x)\log\beta(x) = -[x\log x+(1-x)\log(1-x)], \\ \label{eq:b1x}
    B_1(x) &= 1-c_1(x)-c_2(x) = -2, \\ \label{eq:d1x}
    D_1(x) &= (1-c_1(x)-c_2(x))(\gamma+1) + \frac{5}{2} + c_1(x)\log\alpha(x) + c_2(x)\log\beta(x) = \frac{1}{2} - 2\gamma -\log x -2\log(1-x).
\end{align}
For computing $\mathcal{Q}_N(x,1) = -\left.\partial_p \mathcal{W}_N(x,p)\right|_{p=1}$, we may take the $p\to 1$ limit of $\mathcal{W}_N(x,p)$ near $p=1$. Defining $p=1+\delta$ and using equations~\eqref{eq:apx}-\eqref{eq:dpx}, we obtain that
\begin{align}
    A_{1+\delta}(x) &= \frac{1 - x^{1-\delta} - (1-x)^{1-\delta}}{-\delta(1-\delta)} \\ \notag
    &= -[x\log x + (1-x)\log(1-x)] + \delta\left\{ \frac{x\log^2x + (1-x)\log^2(1-x)}{2} - x\log x - (1-x)\log(1-x) \right\} + \mathcal{O}[\delta^2], \\
    B_{1+\delta}(x) &= \frac{1 - 2(1-x)^{-\delta} - x^{-\delta}}{-\delta} = \frac{2}{\delta} - \log x - 2\log(1-x) + \frac{\delta}{2}\left[ \log^2x + 2\log^2(1-x) \right] + \mathcal{O}[\delta^2], \\
    D_{1+\delta}(x) &= 1 + \zeta(\delta) -2\zeta(1+\delta) = -\frac{2}{\delta} + \frac{1}{2} -2\gamma + \delta\left( \zeta'(0) + 2\gamma_{(1)} \right) + \mathcal{O}[\delta^2],
\end{align}
with $\gamma_{(m)}$ the $m$th Stieltjes constant with $\gamma_{(0)} = \gamma$ and $\gamma_{(1)}\approx -0.0728158$. Since $N^{-\delta} = 1 - \delta\log N + \mathcal{O}[\delta^2]$,  
we get%
\begin{equation}
\begin{split}
    \mathcal{W}_N(x, 1+\delta) &= -[x\log x + (1-x)\log(1-x)]N -2\log N + \frac{1}{2} - 2\gamma -[\log x + 2\log(1-x)] \\
    &+\delta \bigg\{ \left[\frac{x\log^2x + (1-x)\log^2(1-x)}{2} - x\log x - (1-x)\log(1-x) \right]N \\
    &\qquad\qquad+ [x\log x+ (1-x)\log(1-x)]N\log N + \log^2 N + [\log x + 2\log(1-x)] \log N \\
    &\qquad\qquad+ \frac{x\log x + (1-x)\log(1-x)}{2} + \zeta'(0) + 2\gamma_{(1)} \bigg\} + \mathcal{O}[\delta^2].
\end{split}
\end{equation}
To obtain $\mathcal{Q}_N(x,1)$, note that $\mathcal{Q}_N(x,1) = \lim_{p\to 1}[-\partial_p\mathcal{W}_N(x,p)] = \lim_{\delta\to 0} [-\partial_\delta\mathcal{W}_N(x,1+\delta)]$, so finally
\begin{align}\notag
    \mathcal{Q}_N(x,1) &= \left[ x\log x + (1-x)\log(1-x) - \frac{x\log^2x + (1-x)\log^2(1-x)}{2} \right] N  - [x\log x+ (1-x)\log(1-x)]N\log N \\
    &- \log^2 N - [\log x + 2\log(1-x)] \log N - \frac{x\log x + (1-x)\log(1-x)}{2} - \zeta'(0) - 2\gamma_{(1)}.
\end{align}

For $p=2$, \eqlabel{eq:apx} presents another pole. Using again~\eqlabel{eq:zeta1}, we have that
\begin{equation}
\begin{split}
    \mathcal{W}_N(x,2) &= (1-\alpha(x)-\beta(x))\zeta(2)N + \log N + (1-c_1(x)-c_2(x))\zeta(2) + \log\alpha(x) + \log\beta(x) + 2 + \gamma + \mathcal{O}[N^{-1}] \\
    &= \log N + \log x(1-x) + 2 + \gamma -2\zeta(2) + \mathcal{O}[N^{-1}].
\end{split}
\end{equation}
Using $p=2+\delta$, we can obtain $\mathcal{Q}_N(x,2)$ in a similar way as for $p=1$. Now we need the expansions of $\zeta_z(1+\delta)$ and $\zeta_z(2+\delta)$, where $\zeta_z(s+\delta) = \zeta_z(s) + \delta \!\left.\partial_t\zeta_z(t)\right|_{t=s} + \mathcal{O}[\delta^2]$ where, for large $z$, we have as first derivatives
\begin{align}
    \left.\partial_t \zeta_{Nc+d}(t)\right|_{t=1} = -\frac{1}{2}(\log N + \log c)^2 - \gamma_{(1)} + \mathcal{O}\!\left[ \frac{\log N}{N} \right]
    \quad\text{and}\quad\left.\partial_t \zeta_{Nc+d}(t)\right|_{t=2} = \zeta'(2) + \mathcal{O}\!\left[ \frac{\log N}{N} \right].
\end{align}
Substituting into~\eqlabel{eq:wnxp}, we obtain that
\begin{align}
    \mathcal{W}_N(x,2+\delta) &= \mathcal{W}_N(x,2) \\ \notag &- \delta \underbrace{\bigg[ \frac{\log^2N + 2\log N\log x(1-x) + \log^2x + \log^2(1-x)}{2} + \log Nx(1-x)  + 2\zeta'(2) + \gamma_{(1)} + 1 \bigg]}_{=\mathcal{Q}_N(x,2)} + \mathcal{O}[\delta^2].
\end{align}

Having addressed $\mathcal{W}_N(x,p)$ and $\mathcal{Q}_N(x,p)$ in the large-$N$ limit for arbitrary $p$, now we can insert these expressions into the participation ratio definition. In this limit, the participation ratio becomes
\begin{equation}\label{eq:eta_largeN}
    \lim_{N\to\infty} \mathbb{E}[\eta_N(p)] = \lim_{N\to\infty} \frac{8\eta_\text{crit}(N-2)}{N_{\max}}\int_0^1 \text{d}x \mathcal{W}_N(x,p)\exp\left[ p\frac{\mathcal{Q}_N(x,p)}{\mathcal{W}_N(x,p)} \right] = \eta_\text{crit} I(p),
\end{equation}
where we define $I(p) = \lim_{N\to\infty} I_N(p)$ by means of the integral
\begin{equation}\label{eq:ip}
    I_N(p) = \frac{6}{N^2} \int_0^1 \text{d}x \mathcal{W}_N(x,p)\exp\left[ p\frac{\mathcal{Q}_N(x,p)}{\mathcal{W}_N(x,p)} \right]. 
\end{equation}

Using equations~\eqref{eq:apx} and~\eqref{eq:a1x}, we obtain that for $0\le p<2$,
\begin{equation}
    \frac{\mathcal{Q}_N(x,p)}{\mathcal{W}_N(x,p)} = \log N -\frac{\partial_p A_p(x)}{A_p(x)} + \mathcal{O}\!\left[ \frac{\log N}{N} \right] \implies I(p) = \lim_{N\to\infty} I_N(p) = 6 \int_0^1 \text{d}x A_p(x)\exp\left[ -p\frac{\partial_p A_p(x)}{A_p(x)} \right],
\end{equation}
which is constant. This integral satisfies that $0< I(p)\le 1$, with equality for $p=0$ and that $I(p)$ is monotonically decreasing with increasing $p\in[0,2)$ (see~\figlabel{fig:ip}), thus fewer terms participates. For $p=2$, we have instead that
\begin{equation}
    \frac{\mathcal{Q}_N(x,2)}{\mathcal{W}_N(x,2)} = \frac{1}{2} \left( \log N + \log x(1-x) \right) - \frac{\gamma}{2} + \zeta(2) + \mathcal{O}\!\left[\frac{1}{\log N}\right] \implies \lim_{N\to\infty} \frac{N I_N(2)}{\log N} = e^{2\zeta(2)-\gamma} \approx 15.0692.
\end{equation}
For $p>2$, the leading term in $\mathcal{W}_N(x,p)$ is instead the constant contribution $D_p(x)$ [equations~\eqref{eq:dpx} and~\eqref{eq:d1x}], so
\begin{equation}
    \lim_{N\to\infty} N^2I_N(p) = 6 \int_0^1 \text{d}x D_p(x)\exp\left[ -p\frac{\partial_p D_p(x)}{D_p(x)} \right] = \text{const.} \implies \mathbb{E}[\eta_N(p)] \propto \frac{1}{N^2}.
\end{equation}
\begin{figure}[!tb]
    \centering
    \includegraphics[width=.55\linewidth]{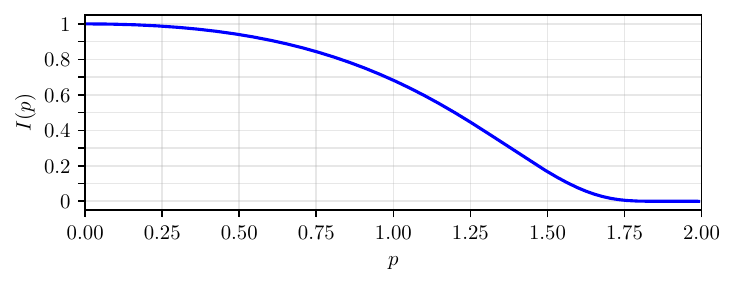}\vspace{-1mm}%
    \caption{\textbf{Evaluation of $\bm{I(p)}$.}  Numerical computation of $I(p) = \lim_{N\to\infty} I_N(p)$ [\eqlabel{eq:ip}] across $p\in[0,2)$.}\label{fig:ip}
\end{figure}%

Therefore, we identify three different regimes according to the asymptotic behaviour displayed in the thermodynamic limit: i) a constant behaviour controlled by $p\in[0,2)$, thus a superextensive scaling might be obtained, ii) a smooth $\log N/N$ decay for $p=2$, and iii) a quadratic decay with increasing system size for $p>2$, reminiscent of extensive scaling performance. In~\figlabel{fig:eta_plaw} we compute for $N\in[2,144]$ the expected participation ratios for $p\in[0,4]$ via scrambled quasi-Monte Carlo evaluation of equation X including all $C_i$ contributions with the same configuration as in~\figlabel{fig:participation_exact}. Moreover, we include the same computation but including only the $C_2$-family related contributions and their corresponding asymptotic behaviour as a reference. We observe that the predicted asymptotic behaviours match the obtained numerical results. As a side note, the $N=144$ upper bound constraint is due to the numerical limitations present in the quasi-Monte Carlo implementation used~\cite{qmc}. It is limited to $21201$ variables, while our integral requires for the $k$th battery cell $4N + 4(k-1)(N-k)$, achieving its maximum at $k=\lfloor N/2 \rfloor$ and leading to $N=144$ as the largest available system size using this method.
\begin{figure}[!tb]
    \centering
    \includegraphics[width=\linewidth]{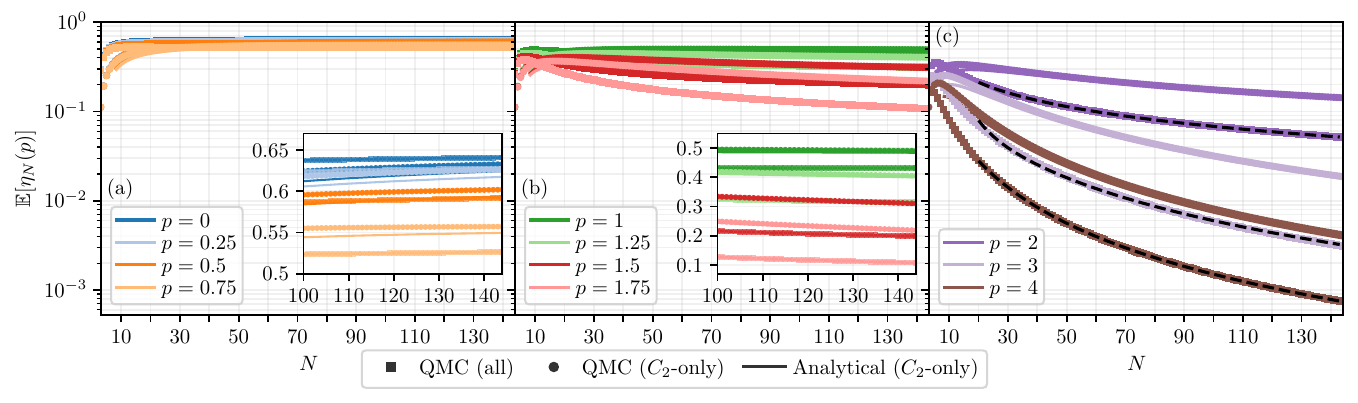}\vspace{-2mm}%
    \caption{\textbf{Power-law locality-decaying participation ratios.} Using the same configuration as in~\figlabel{fig:participation_exact}, equation X is evaluated for $N\in[3,144]$ and $p\in[0,4]$ considering all $C_i$ families (square) and only the $C_2$ family (triangle) contributions. We include the predicted asymptotic behaviours (solid line) for the latter and fit data points in $N\in[30, 144]$ to $a\log N/N$ for $p=2$ and $a/N^2$ for $p=3,4$ (dashed) as a reference.
    }\label{fig:eta_plaw}\vspace{-2mm}%
\end{figure}%

For completeness, we compare these regimes with those obtained when the battery-charger commutator is bounded as $\norm{[H_\text{b}, H_\text{c}]} \le \widetilde{\mathcal{U}}_p(N)$ using the $\norm{C_i}$ upper bounds derived in~\seclabel{sec:upper_bound} and~\eqlabel{eq:plaw_renormalization}. Following the same procedure as before, the asymptotic leading terms are
\begin{equation}
    \widetilde{\mathcal{U}}_p(N) \sim \begin{cases}
        N^2 & \text{if }0\le p< 1/2 \\
        N^2/\sqrt{\log N} & \text{if }p=1/2 \\
        N^{5/2-p} & \text{if }1/2 < p < 3/2 \\
        N\log N & \text{if }p=3/2 \\
        N & \text{if }p>3/2
    \end{cases}.
\end{equation}
Consequently, the bound is superextensive for $p\le 3/2$, which marginally scales as $N\log N$ for $p=3/2$, and extensively for $p>3/2$.

\bibliography{bibfile}